\documentclass[twocolumn,9pt]{article}
\usepackage[square,sort,comma,numbers]{natbib}
\usepackage[utf8]{inputenc}
\usepackage[margin=0.9in,paper=letterpaper]{geometry}

\usepackage{paralist} % For compact itemize and enumerate
\usepackage{outlines}
\usepackage{multicol}
\usepackage{booktabs} % For formal tables
\usepackage{amssymb}  % For bullet in itemize
\usepackage{graphicx} % For figures
\usepackage{makecell}
\usepackage{wrapfig}
\usepackage{lscape}
\usepackage{appendix}
\usepackage[most]{tcolorbox} % For border for figures
\usepackage{subfigure} % For subfigures
\usepackage{tikz,pgfplots,pgfplotstable} % CMU style TiKz plots
\usepackage{mdframed} % Infobox at the top of the document
\pgfplotsset{compat=1.7} % CMU style TiKz plots
\usepackage{pifont} % Symbol fonts
\usepackage{amsmath} % Math symbols
\usepackage{bm} % Bold Math symbols
\usepackage{multirow} % Vertical text in a table
\usepackage{enumitem} % Required for fancy itemize / enumerate

\usepackage{url}

\usepackage{libertine}
\usepackage{libertinust1math}
\usepackage[T1]{fontenc}

\begin{document}

%\pagestyle{fancy}
%\cfoot{\thepage}
\title{Double Patterns: A Usable Solution to Increase the Security of Android Unlock Patterns\thanks{A version of this paper is appears at the 2020 Annual Computer Security Applications Conference (ACSAC'20).}}


% \author{Timothy J. Forman}
% \affiliation{%
%   \institution{U.S. Naval Academy}}
% \email{tforman37@gmail.com}

% \author{Adam J. Aviv}
% \affiliation{%
%   \institution{The George Washington University}}
% \email{aaviv@gwu.edu}

\author{Timothy J. Forman\\
{\small U.S. Naval Academy}\\
{\small \texttt{tforman37@gmail.com}}
\and
Adam J. Aviv\\
{\small The George Washington University}\\
{\small \texttt{aaviv@gwu.edu}}}

\date{}
%\renewcommand{\shortauthors}{Timothy J. Forman and Adam J. Aviv}


% <concept_id>10002978.10002991.10002992.10011618</concept_id>
% <concept_desc>Security and privacy~Graphical / visual passwords</concept_desc>
% <concept_id>10003120.10003121</concept_id>
% <concept_desc>Human-centered computing~Human computer interaction (HCI)</concept_desc>
% <concept_id>10003120.10003121.10003122.10010854</concept_id>
% <concept_desc>Human-centered computing~Usability testing</concept_desc>

% \ccsdesc[500]{Security and privacy~Graphical / visual passwords}
% \ccsdesc[500]{Human-centered computing~Human computer interaction (HCI)}
% \ccsdesc[500]{Human-centered computing~Usability testing}


\maketitle
\begin{abstract}
    Android unlock patterns remain quite common. Our study, as well as others, finds that roughly 25\% of respondents use a pattern when unlocking their phone. Despite known security issues, the design of the pattern interface remains unchanged since first launch. We propose Double Patterns, a natural and easily adoptable advancement on Android unlock patterns that maintains the core design features, but instead of selecting a single pattern, a user selects two, concurrent Android unlock patterns entered one-after-the-other super-imposed on the same 3x3 grid. We evaluated Double Patterns for both security and usability by conducting an online study with $n=634$ participants in three treatments: a control treatment, a first pattern entry blocklist, and a blocklist for both patterns. We find that in all settings, user chosen Double Patterns are more secure than traditional patterns based on standard guessability metrics, more similar to that of 4-/6-digit PINs, and even more difficult to guess for a simulated attacker.  Users express positive sentiments in qualitative feedback, particularly those who currently (or previously) used Android unlock patterns, and overall, participants found the Double Pattern interface quite usable, with high recall retention and comparable entry times to traditional patterns. In particular, current Android pattern users, the target population for Double Patterns, reported SUS scores in the 80th percentile and high perceptions of security and usability in responses to open- and closed-questions. Based on these findings, we would recommend adding Double Patterns as an advancement to Android patterns, much like allowing for added PIN length.
\end{abstract}
%%% Local Variables:
%%% mode: latex
%%% TeX-master: "main"
%%% End:

%\keywords{Double Pattern, Android Unlock Pattern, Biometrics, Mobile Authentication, Security, Usability}

\section{Introduction}
\label{sec:intro}
There are two primary, knowledge based authentication methods for unlocking mobile devices: 4/6-digit PINs and Android unlock patterns. Prior work has shown (and is confirmed herein) that patterns are used by about 25\% of users~\cite{markert-20-pin-blocklist,aviv2015bigger,harbach2014sa}. Despite there being 389,112 possible patterns (more than 38x more than 4-digit PINs), it is known that users likely select from a much smaller subset of patterns in easily predictable  ways~\cite{uellenbeck-13-pattern,aviv2015bigger}, much more so than how users select 4-/6-digit PINs~\cite{markert-20-pin-blocklist,wang-17-pin}.

Unlike the shift from 4-digit to 6-digit PINs (or longer), there has not been a significant change to the interface of Android patterns since first launched on the Android T-Mobile G1 (or HTC Dream) in 2008 as the first commercially available Android device\footnote{\url{https://en.wikipedia.org/wiki/HTC_Dream} (viewed on \today)}, where a pattern was the only available unlock authentication option. There have been various proposals to improve patterns, including providing user guided selection~\cite{cho2017syspal}, rearrangement of the contact points~\cite{tupsamudre2017PassO}, strength meters~\cite{song-15-pattern-psm,andriotis-14-pattern-psm,sun-14-pattern-psm}, and expansion to a 4x4 grid~\cite{aviv2015bigger}. These proposals require either a departure from the distinctly simple selection interface or additional interventions that may frustrate users, driving them away from selecting their preferred patterns. More natural expansions of patterns, such as 4x4 patterns, have unfortunately been shown to not increase security~\cite{aviv2015bigger} against a throttled attacker making a limited number of guesses. 

\begin{figure}[t]
\centering
\fbox{\includegraphics[width=.3\linewidth]{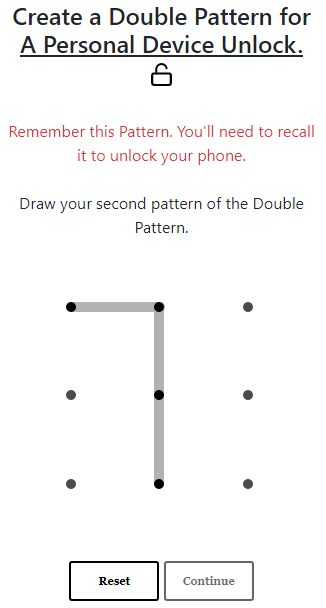}}
\hspace{1em}
\fbox{\includegraphics[width=.3\linewidth]{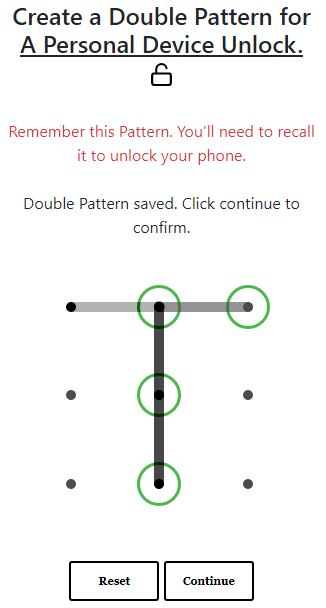}}
\caption{Double Pattern Creation Process}
\label{fig:doublepattern}
\end{figure}

To address these challenges, we offer a novel improvement to Android patterns: {\em Double Patterns (or DPatts)}, whereby a user selects two sequential, superimposed patterns as their unlock authentication (see Figure~\ref{fig:doublepattern}). Utilizing an identical 3x3 grid, the user draws their first pattern, lifts, and then draws their second pattern, with both patterns being displayed at the same time. This provides an increase in the  visual complexities for users to select patterns and a large increase in the total number of DPatts (151,407,759,432 options) as compared to traditional unlock patterns. The design of DPatt takes advantage of the popular 3x3 interface and encourages users to select more secure patterns through the natural increased complexity of multiple patterns. 

We conducted an online survey on Amazon Mechanical Turk to assess the potential usability and security of DPatt, first in a preliminary survey with $n=286$ participants and then in a main study with $n=634$. In the course of the survey, participants both selected a DPatt and answered questions about their experiences and perceptions of DPatt as an alternative for Android patterns. We considered three treatments for DPatt: a control treatment and two blocklist enforced treatments. Using the preliminary survey data, we developed two blocklists, one in which the first pattern of a DPatt was blocklisted and must be re-selected before selecting the second pattern, and the second blocklist blocked a set of common DPatts.

We evaluated the security of the DPatts using guessability metrics and compared the results to the security of 4/6-digit PINs and traditional Patterns (without a blocklist). When the attacker is throttled, or limited in the number of guesses, we find that DPatt's security metrics are more in-line with (but still weaker than) 4-/6-digit PINs. After 30 guesses, a perfect knowledge attacker~\cite{bonneau2012science} would only guess 28\% of DPatts compared to 35\% of patterns, and only 20\% of 4-/6-digit PINs are guessed after 30 attempts. 
However, when considering a simulated attacker that guesses an unknown set of DPatts based on modeling from a sample set, DPatts outperform both traditional patterns and 4-/6-digit PINs. After 30 guesses an attacker would only guess 5.3\% of DPatts compared to 23.6\% of patterns, 7.6\% of 4-digit PINS, and 9.0\% of 6-digit PINs. 
The addition of either blocklist (first pattern or Double Pattern) also greatly improved the security metrics for both a perfect knowledge (18\% and 20\% after 30 attempts, respectively) and simulated attacker (1.9\% and 0.9\% after 30 attempts, respectively).

Participants recalled their selected DPatts at very high rates (97\% success after 1.3 attempts), with extremely comparable entry speeds per attempt of 3.35s (on average across treatments). Prior work suggests that traditional Android pattern entry takes 3.0s and 4-digit PINs take 4.7s~\cite{harbach2014sa}. 
Among the 25\% of participants currently using patterns as their method to unlock their device, they reported System Usable Scale (SUS) scores of $78.27$, placing it in the 80-84th percentile range. Even participants who currently do not utilize a pattern report good and acceptable SUS scores of $71.47$, falling in the 60-64th percentile range. This suggests that if deployed to the target audience of current pattern users, they would be open to moving towards Double Patterns from the traditional pattern. 

This paper makes the following contributions:
\begin{itemize}
    \item We propose a natural extension to Android unlock patterns, Double Patterns (DPatts), where users must enter two superimposed patterns, in sequence, as their authentication.
    \item We show that DPatts significantly improve the security of patterns using guessability metrics, for both a perfect-knowledge and simulated attacker, and may be more secure than 4-/6-digit PINs against simulated attackers. 
    \item The usability of DPatts is not degraded by requiring multiple pattern entries, with per-attempt entry speeds comparable to traditional pattern entry and high short-term recall rates.
    \item Participants reported usability as good and acceptable and had a high perception of security regarding DPatt, which would encourage adoption. 
    \item Participants currently utilizing patterns, the target population for DPatt, reported even higher positive sentiment for DPatt, both in usability metrics and perceived security of DPatt authentication.
\end{itemize}

Our results suggest that Double Patterns are an extremely viable improvement to the traditional Android unlock pattern, both from a security and usability perspective, and that current Android pattern users would be willing to adopt DPatt as a natural extension to Android patterns. 

%%% Local Variables:
%%% mode: latex
%%% TeX-master: "main"
%%% End:

\section{Double Patterns}
\label{sec:dpatt}
Double patterns are built upon Android unlock patterns, which are a knowledge-based authentication system, whereby a user must recall a pre-selected ``pattern'' by connecting points on a 3x3 grid, without lifting. For example, the {\em left} side of Figure~\ref{fig:doublepattern} shows a traditional pattern, which could have been entered by selecting the top-left point and tracing downward, or selecting the bottom-middle point and tracing upwards (two different patterns). A pattern must be drawn such that at least four contact points are used, no point is used more than once, and any unselected point cannot be avoided or traced over without being previously selected. In total there are 389,112 possible patterns~\cite{aviv-10-smudge}.

Double Patterns (or DPatts) are also designed to be a knowledge based authentication system whereby a participant must recall {\em two} previously selected Android patterns entered in sequence. Both patterns in a DPatt are superimposed, allowing for more complex visual shapes, and the patterns must also be entered in the exact order to be authenticated. For example, Figure~\ref{fig:doublepattern} the two inverted `L' patterns combine to form a `T' pattern as the DPatt.

The same restrictions on the individual Android patterns exist---at least four points, a point cannot be used more than once, and unselected points cannot be avoided---but after entry of the first pattern, all contact points can be used in the second pattern, as if it was drawn independently. The only restriction on the second pattern is that it {\em must} be a different pattern than the first. There are 151,407,759,432 total DPatts. 

%%% Local Variables:
%%% mode: latex
%%% TeX-master: "main"
%%% End:

\section{Related Works}
\label{sec:related}
There is much prior work on Android patterns. Andriotis et al.~\cite{andriotis-13-pilot-study} provided one of the first studies regarding user habits when selecting unlock patterns. Uellenbeck et. al~\cite{uellenbeck-13-pattern} collected a sample of Android patterns and analyzed the guessability of so called ``defensive'' patterns chosen to purposely avoid guessing and ``offensive'' patterns chosen to guess others' patterns. Uellenbeck et. al found that Android patterns, theoretically, were only as diverse as selecting a random 3-digit PIN. Aviv et al.~\cite{aviv2015bigger} conducted an online study asking participants to self-report patterns, confirming that user selection of Android patterns are less diverse than other authentication choices. Loge et al.~\cite{loge-16-pattern-user-choice} investigated selection of patterns in different settings, such as to secure a banking app or shopping cart in addition to phone unlocking, finding, again, that the security of Android patterns is challenged. A summary and comparison of these results and others related to Android pattern is provided by Aviv and D\"urmuth~\cite{avivSurvey}.

Android patterns have also been the subject of attack. Aviv et al. first demonstrated a smudge-based side channel attack~\cite{aviv-10-smudge}, whereby residues left on the smartphone screen reveal prior pattern entries, and have since been shown to boost guessing performance of an attacker~\cite{cha-17-pattern-smudges}. Patterns have also been shown to be less resilient to shoulder surfing~\cite{schaub-12-passwords-on-smartphones, aviv-17-shoulder-surfing-baseline}, as well as video-based reconstruction attacks~\cite{guixin2018video}. Even the onboard sensors of smartphones can reveal information about pattern input~\cite{aviv-12-practicality}.

Due to the insecurity and lack of diverse choices for Android patterns, there have been many proposals for improvement. This included modifications to avoid shoulder surfing~\cite{zezschwitz-15-pattern-shoulder-surfing, de-luca-14-shoulder-surfing} and smudge attacks~\cite{kwon2014tinylock, schneegass2014smudgesafe}, which maintain the primary design of Patterns but transform the input procedure. Other more radical proposals include rearranging the points of the pattern, such as into a ring~\cite{tupsamudre2017PassO}.

Password meters are another common proposal for improving pattern selection. Andriotis et al.~\cite{andriotis-14-pattern-psm}, Sun et al.~\cite{sun-14-pattern-psm}, and Song et al.~\cite{song-15-pattern-psm} each proposed visual based strength metrics and a display meter to boost diversity of patterns selected. While meters may be an effective means of changing behavior, Golla et al. demonstrated that strength metrics used in these meters do not correlate with security, and likely, just the presence of the meter changes behavior~\cite{golla2019inaccuracy}.

Cho et al. proposed SysPal~\cite{cho2017syspal}, which highlights certain contact points that {\em must} be used as part of the pattern, restricting users to select different patterns but also more diverse ones. von Zezschwitz et al. suggested that background images can improve pattern selection, if sufficiently complex~\cite{zezchwitz2016onquant}. Aviv et al. investigated 4x4 patterns~\cite{aviv2015bigger}, finding that there are little benefits from larger patterns.

Double patterns offers a new direction in improving Android patterns as it is a natural and straightforward progression in design. DPatts maintain the same popular interface and improve security without direct interventions, such as highlighting points, providing background images, or including password meters. Additionally, the use of multiple patterns increases the burden on observation attacks whereby shoulder surfing and video-based attacks would be more challenging due to the added complexity.

In evaluating the performance of DPatts we also consider research into other mobile unlock authentications, such as 4-/6-digit PINs. Bonneau et al. studied user choice of 4-digit PINs in the credit-card, chip-and-pin system~\cite{bonneau2012birthday}, finding that many users select PINs derived from dates. Wang et al. studied 4-/6-digit PINs derived from leaked password data sets~\cite{wang-17-pin}, finding subtle differences between English speaking and Chinese speaking users' selection of PINs, and that the advantages of 6-digit PINs is minimal. Markert et al. collected 4-/6-digit PINs in the context of smartphone unlock, further demonstrating that there are minimal benefits of 6-digit PINs and the current use of blocklists~\cite{markert-20-pin-blocklist}.

We compare DPatts to the security of 4-/6-digit PINs based on data provided by Markert et al. as it is specifically primed for smartphone authentication. We use data from von Zezchwitz et al.~\cite{zezchwitz2016onquant}, Aviv et al.~\cite{aviv2015bigger}, Uellenbeck et al.~\cite{uellenbeck-13-pattern}, and Loge et al.~\cite{loge-16-pattern-user-choice} to compare DPatt to traditional patterns, as well as to derive a synthetic DPatt data set used in our guessing analysis.

%%% Local Variables:
%%% mode: latex
%%% TeX-master: "main"
%%% End:

\section{Methodology}
\label{sec:methods}

To evaluate Double Patterns (DPatts), we developed an online, browser-based survey and recruited via Amazon Mechanical Turk (MTurk).  Using their own personal mobile devices, participants completed the survey by creating/recalling a DPatt, as well as answering questions about their experience. For the main study, we recruited $n=634$ participants in three treatments: a control treatment and two blocklist treatments. The complete survey material can be found in Appendix~\ref{app:survey}.

\begin{table}[t]
    \centering
    \caption{Participant Device Utilization}
      \label{tab:dev} 

    \resizebox{.9\columnwidth}{!}{%
    \begin{tabular}{ r | c c c | c}
    & Control & BL-First & BL-Both & \textbf{Overall} \\
    \toprule
    Iris Recognition & 0 & 2 & 1 & 3 (.5\%) \\
    Finger Print & 108 & 106 & 111 & 325 (51.3\%) \\
    Facial Recognition & 26 & 26 & 26 & 78 (12.3\%) \\
    No Biometric & 72 & 70 & 67 & 209 (33.0\%) \\
    Other Form & 3 & 7 & 6 & 16 (2.5\%) \\
    \midrule
    Pattern & 57 & 49 & 56 & 162 (25.6\%)\\
    4-Digit PIN & 96 & 89 & 98 & 283 (44.6\%) \\
    6-Digit PIN & 29 & 34 & 36 & 99 (15.6\%) \\
    PIN of Other Length & 8 & 8 & 7 & 23 (3.6\%) \\
    Alpha-Numeric & 6 & 12 & 9 & 27 (4.3\%) \\
    Not Listed & 11 & 16 & 7 & 34 (5.4\%) \\
    Prefer not say & 2 & 3 & 1 & 6 (0.9\%) \\
    \bottomrule
    \textbf{Total} & {\bf 209} & {\bf 211} & {\bf 214} & \textbf{634} \\ 
    \end{tabular}
    }
\end{table}

%%% Local Variables:
%%% mode: latex
%%% TeX-master: "../main"
%%% End:

\subsection{Survey Outline}
There are 12 sections to the survey. Participants are first informed about DPatts and allowed to practice creating DPatts, before being tasked to select one that they might use to unlock their smartphone. Following, participants answer questions about their experience selecting a DPatt and the perceived usability and security, before being asked to recall their selected DPatt. It took participants, on average, 7.3 minutes to complete the survey; the survey, in its entirety, can be found in Appendix~\ref{app:survey}. The protocol for the study was approved by the IRB of our institution(s). 
\begin{enumerate}
\item {\em Purpose of Study/Informed Consent}: Participants were informed about (and consented to) the study before proceeding, this included details that participants would be asked to select/recall a Double Pattern and answer questions about their experiences.
\item {\em Device Usage:} Participants were asked to provide background on the number of mobile devices (e.g., smartphones) they currently use, as well as which mobile authentication method they currently use to unlock their devices. For those that indicated that they use a biometric, we provided a follow up question asking how the participant unlocks their device following a phone reset, or when their biometric fails. Details of the device usage can be found in Table~\ref{tab:dev}, and just over 25\% of our participants use a Android pattern to secure their smartphone.
\item {\em Android Patterns/Double Patterns Background:} As not all participants were familiar with Android patterns, we provide information about them as well as how they related to Double Patterns, such as:``Double Pattern Locks are the same as Pattern Locks but require you to `draw' two shapes on the same 3x3 grid of contact points. The combination of the two patterns entered in the same order is now used to unlock your smartphone.''
\item {\em Practice:} Participants were presented with the DPatt interface and asked to practice using it by creating (and confirming) a DPatt before proceeding. This provided familiarity and ensured that participants' first interaction would not be used as part of the analysis.
\item {\em Instructions:} Now familiar, participants were informed that the next DPatt selected would be used as part of the study. They were asked to ``create a Double Pattern you would likely use for a personal device unlock, such as you would use on your smartphone.'' They were also instructed that ``you will need to recall this Double Pattern later in the survey, so choose something that is secure and memorable.'' Two confirmations were requested at this point: (a) that the participant understood that they are supposed to create a Double Pattern for a personal device unlock, and (b) that they should not write down their Double Pattern or use other aids to help them remember it. 
\item {\em Selection:} Participants were then instructed to select a DPatt, and the instructions to select something for unlocking their smartphone were also included on this page (see Figure~\ref{fig:doublepattern}). During selection, a participant may experience an enforcing blocklist, which disallows a predetermined set of DPatts. We describe the treatments below in Section~\ref{sec:method:treatments}. 
\item {\em Post-Entry:} After selection participants are asked Likert agreement for two statements: if the DPatt provides adequate security for unlocking their device, and their difficulty in selecting the DPatt. As well, participants were asked an open-text response regarding their strategy in selecting the DPatt they chose.
\item {\em SUS:} The 10 question System Usability Scale was then administered to determine the perceived usability of DPatt.
\item {\em Recall:} After the distractor tasks above, participants were asked to recall their selected DPatt. After five attempts, if they were unable to recall their DPatt, participants were allowed to indicate that they could not remember and proceed with the survey.
\item {\em Security Comparison:} Participants were then asked about the perceived security of DPatt itself, and to compare it to Android patterns, 4-digit PINs, 6-digit PINs, and alpha-numeric passwords using a Likert agreement scale.
\item {\em Use Double Pattern from Survey:} Then, participates could indicate and explain if they would use the DPatt selected in the survey as their own unlock authentication, if Double Patterns were used, or if they would choose a different DPatt (or were unsure.)
\item {\em Demographics:} Finally, participants were asked to provide demographic information, such age, identified gender, dominant hand, education, and technical background.
\end{enumerate}

\begin{figure}[t]
  \centering
  \fbox{\includegraphics[width=0.5\linewidth]{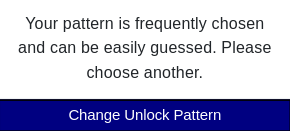}}
  \caption{Blocklist Warning}
  \label{fig:bl_message}
\end{figure}
%%% Local Variables:
%%% mode: latex
%%% TeX-master: "../main"
%%% End:

\begin{table}[t]
      \caption{Results of Asking Participants if they were comfortable using the DPatt they selected.}
    \label{tab:would_use}
    \centering
    \resizebox{\linewidth}{!}{
    \begin{tabular}{c | c c c | c c c | c c c}
    & \multicolumn{3}{c|}{\bf Pattern Users} & \multicolumn{3}{c|}{\bf Non-pattern Users} & \multicolumn{3}{c}{\bf Overall} \\ 
    & {\em Yes} & {\em No} & {\em Unsure} & {\em Yes} & {\em No} & {\em Unsure} & {\em Yes} & {\em No} & {\em Unsure} \\
    \toprule
    Control & 17 & 20 & 20 & 69 & 40 & 43 & 86 & 60 & 63 \\
    BL-First & 24 & 13 & 12 & 60 & 51 & 51 & 84 & 64 & 63 \\
    BL-Both & 26 & 19 & 11 & 72 & 48 & 38 & 98 & 67 & 49 \\
    \bottomrule
    {\bf Total} & {\bf 67} & {\bf 52} & {\bf 43} & {\bf 201} & {\bf 139} & {\bf 132} & {\bf 268} & {\bf 191} & {\bf 175} \\
    \end{tabular}}
\end{table}

%%% Local Variables:
%%% mode: latex
%%% TeX-master: "../main"
%%% End:

\subsection{Treatments}
\label{sec:method:treatments}
Each participant was randomly assigned a treatment:
\begin{itemize}
\item {\bf Control:} In the {\em control} treatment participants received no intervention when selecting/recalling a Double Pattern.
\item {\bf BL-First:} In the {\em blocklist first} (or {\em BL-First}) treatment, a blocklist of first component patterns (the first of the two patterns in a DPatt)  was used to restrict the choices of Double Patterns. After entering the first pattern, and before proceeding to select the second pattern, the participant would be prompted to change their pattern if the first pattern is blocklisted, and would need to continue to select first patterns until one is chosen that is not blocklisted, after which they proceed to selecting their second pattern. 
\item {\bf BL-Both:} In the {\em blocklist both} (or {\em BL-Both}) treatment, a blocklist is used to match {\em both} patterns of a DPatt against a blocklist of disallowed Double Patterns. If the participants selection is blocklisted, they are required to select a different DPatt until they select one that is not blocklisted.  
\end{itemize}
To determine the blocklists, we relied on data collected during prototyping of the survey with $n=286$ participants. During the prototype, we asked participants to select two different Double Patterns (572 total), one for two different scenarios as described in Loge et al.~\cite{loge-16-pattern-user-choice}, either a shopping cart, banking account, or mobile unlock. Participants always select a mobile unlock, and then either shopping cart or banking account. 

For the BL-First treatment, we used the top 20 most common first pattern occurrences, and for the BL-Both pattern, we constructed a blocklist from the 20 most common Double Patterns. There is not prior work on blocklist sizes for patterns, and so we focused on a short blocklist rather than an expansive one. The blocklists used can be found in Appendix~\ref{app:blocklist}. A visual for the blocklist message can be found in Figure~\ref{fig:bl_message}, which matches the blocklist message from iOS~\cite{markert-20-pin-blocklist}, modified for patterns. 

The main study differed from the prototype in three main ways. First, as we did not observe major differences between the scenarios, we elected to have participants focus on just the smartphone unlock scenario, the scenario we envision Double Patterns being deployed. Second, we identified numerous conflations in bias in our questions that were improved in an expanded survey. Third, we implemented two blocklisting treatments alongside our control treatment.

\begin{table}[t]
  \caption{Demographic Information of the Participants}
  \label{tab:demo}
  \footnotesize
  \centering
  \resizebox{\linewidth}{!}{
\begin{tabular}{ r | c c c | c}
 & Control & BL-First & BL-Both & Total \\
\toprule
18-24 & 17 & 26 & 19 & 62 \\
25-29 & 52 & 61 & 55 & 168 \\
30-34 & 45 & 41 & 57 & 143 \\
35-39 & 47 & 35 & 35 & 117 \\
40-44 & 14 & 21 & 17 & 52 \\
45-49 & 17 & 11 & 14 & 42 \\
50-54 & 6 & 7 & 9 & 22 \\
55-59 & 5 & 5 & 3 & 13 \\
60-64 & 2 & 1 & 2 & 5 \\
65+ & 2 & 3 & 3 & 8 \\
Prefer not to say & 2 & 0 & 0 & 2 \\
  \midrule
  Male & 112 & 123 & 135 & 370 \\
  Female & 95 & 84 & 74 & 253 \\
  Non-binary & 0 & 3 & 5 & 8 \\
  Prefer not to say & 2 & 1 & 0 & 3 \\
  \midrule
  Tech & 62 & 66 & 55 & 183 \\
No-Tech & 140 & 139 & 154 & 433 \\
Prefer not to say & 7 & 6 & 5 & 18 \\
    \midrule
High School & 18 & 15 & 25 & 58 \\
Trade & 15 & 6 & 5 & 26 \\
Some-College & 42 & 43 & 47 & 132 \\
Associates & 24 & 27 & 16 & 67 \\
Bachelor's & 82 & 92 & 89 & 263 \\
Master's & 19 & 16 & 26 & 61 \\
Professional & 0 & 5 & 4 & 9 \\
Doctorate & 4 & 3 & 1 & 8 \\
Prefer not to say & 4 & 0 & 0 & 4 \\
\bottomrule
\textbf{Total} & {\bf 209} & {\bf 211} & {\bf 214} & \textbf{634} \\
\end{tabular}}
\end{table}

%appendix
%Combine and shrink
%%% Local Variables:
%%% mode: latex
%%% TeX-master: "../main"
%%% End:

\subsection{Recruitment}
As part of the main study, we recruited $645$ participants on Amazon Mechanical Turk (MTurk), and after removing participants who failed attention checks and/or provided inconsistent responses, we included $n=634$ participants in the our analysis: 209 in the control treatment, 211 in BL-First treatment, and 214 in BL-Both treatment. As is typically the case for MTurk, the sample is mostly young (67.5\% between 25-39), mostly male identifying (58\% male, 40\% female, and 2\% other gender, or prefer not to say), and better educated (75\% with some college or more educational background) than the US as a whole. Participant demographic is presented in Table~\ref{tab:demo}, and additional demographic information can be found in Appendix~\ref{app:demo}.

\subsection{Limitations}
There are a number of limitations with our methods. First, as the survey is online, without direct observations, it is possible for participants to not follow directions properly and provide inconsistent responses. We attempted to mitigate this limitation by including attention tests and reviewing responses. Additionally, collecting data via MTurk introduces some bias in the demographics (as noted above), more balanced collection would be needed to support claims regarding selection for demographics, which we do not make here. As the survey is relatively short, the recall rates of DPatts reflect short-term memorability of DPatts. We believe that high short-term recall would correlate with good long-term recall, but to support stronger claims about memorability, a longitudinal study would be needed.

This survey may be many participants first experience with Android patterns (in any form), and as such the DPatts selected may not fully reflect choices in the wild. To mitigate this, we asked participants if they would use their chosen DPatt on their own device. Overall, 42.3\% said that they would use the DPatt selected during the survey on their device, if Double Pattern was available, 30.1\% expressed they would not, and 27.6\% indicated they were unsure if they would use the DPatt selected during the survey (see Table~\ref{tab:would_use}). The primary reason to not use the DPatt selected during the survey (or were unsure) was the fact that the DPatt was recorded as part of the survey, while a smaller number described wanting something more secure or complex (see Table~\ref{tab:would_use_qual} in Appendix). This suggests that the methods of the survey provide ecological validity for the scope of DPatts users may select in the wild. 

\begin{table*}[!ht]

  \caption{Frequence of Double Patterns}
      \label{tab:freq2}
	\centering
	\resizebox{\textwidth}{!}{
	\begin{tabular}{ccc|c}
	    %\multicolumn{5}{c} {\textbf{Most Frequent Double Patterns}} \\
%	    \hline
          Control & BL-First & BL-Both
          %& Ideal
          & Total \\
        \toprule
        \makecell{ \\ 
        \frame{\includegraphics[width=1cm]{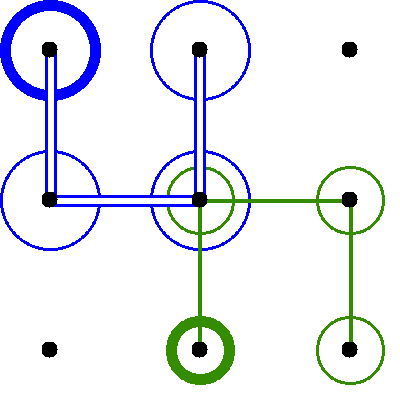}} \\
        \footnotesize{freq = 5} \\ \\
        \frame{\includegraphics[width=1cm]{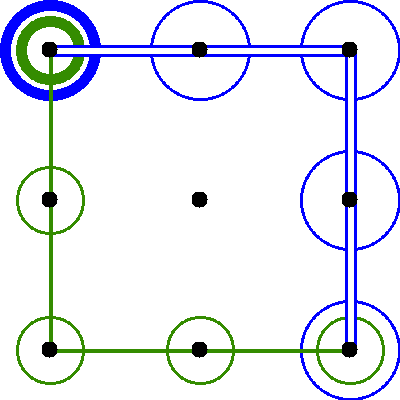}}
        \frame{\includegraphics[width=1cm]{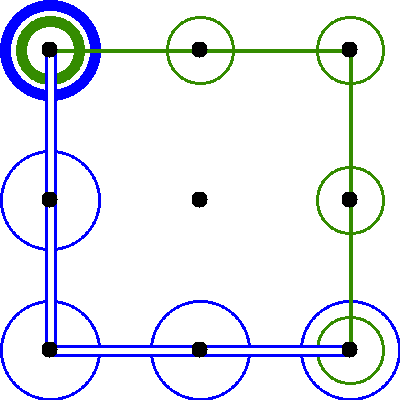}}
        \frame{\includegraphics[width=1cm]{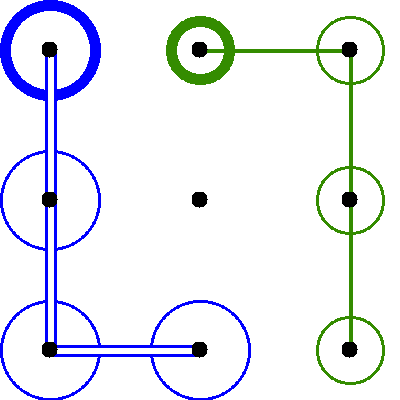}} \\
        \footnotesize{freq = 4} \\ \\
        \frame{\includegraphics[width=1cm]{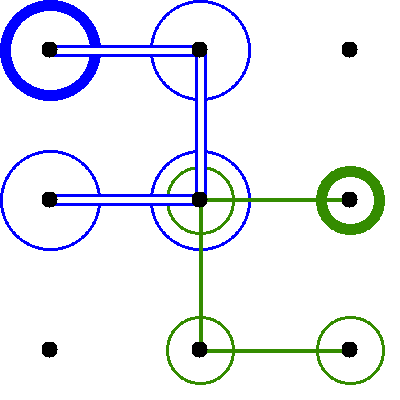}}
        \frame{\includegraphics[width=1cm]{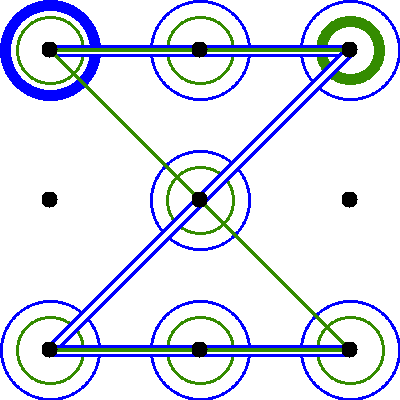}}
        \frame{\includegraphics[width=1cm]{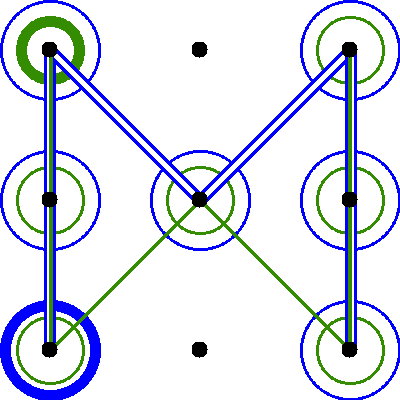}} \\
        \footnotesize{freq = 3} \\
        }&
        \makecell{ \\ 
        \frame{\includegraphics[width=1cm]{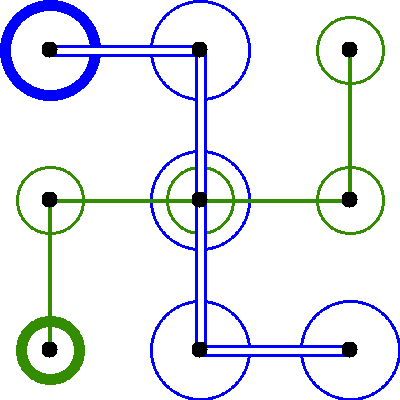}}
        \frame{\includegraphics[width=1cm]{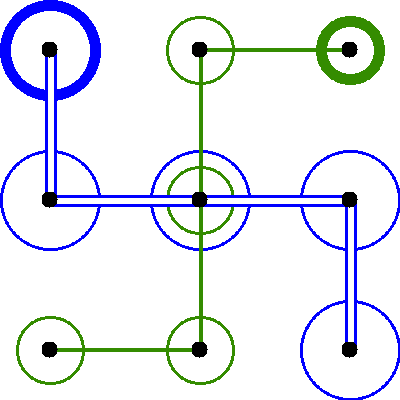}} \\ 
        \frame{\includegraphics[width=1cm]{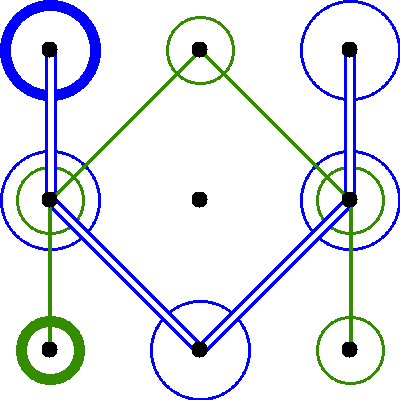}} 
        \frame{\includegraphics[width=1cm]{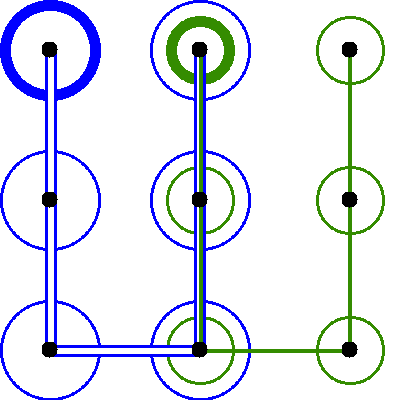}} 
        \frame{\includegraphics[width=1cm]{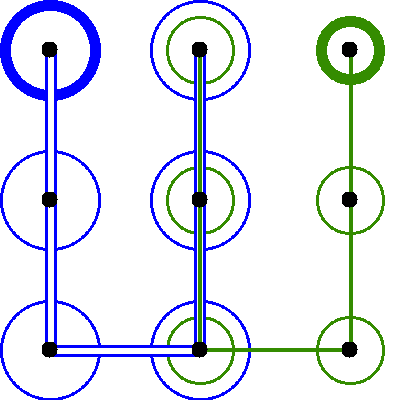}} \\  
        \frame{\includegraphics[width=1cm]{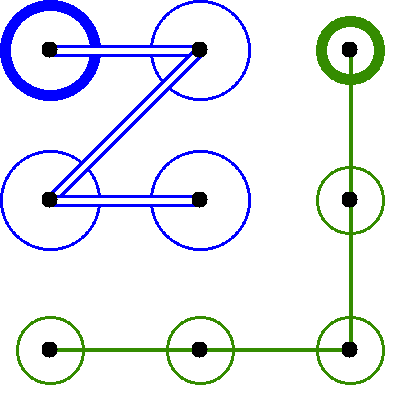}}
        \frame{\includegraphics[width=1cm]{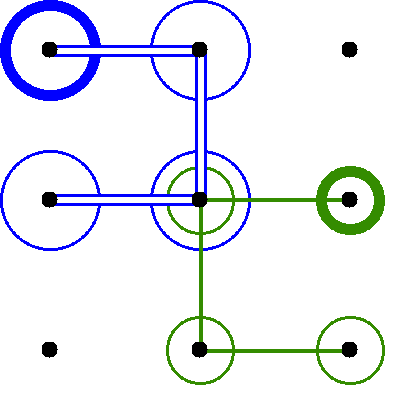}}  
        \frame{\includegraphics[width=1cm]{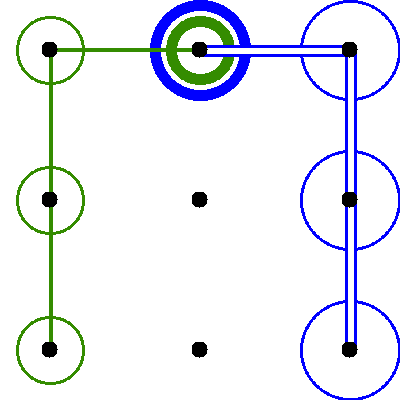}} \\ 
        \footnotesize{freq = 2} \\\\
%        \vspace{2.68cm} \\
        \footnotesize{\emph{Remaining Double Patterns with}} \\
        \footnotesize{\emph{Single Occurrence Omitted}} \\
        }&
        \makecell{ \\ 
        \frame{\includegraphics[width=1cm]{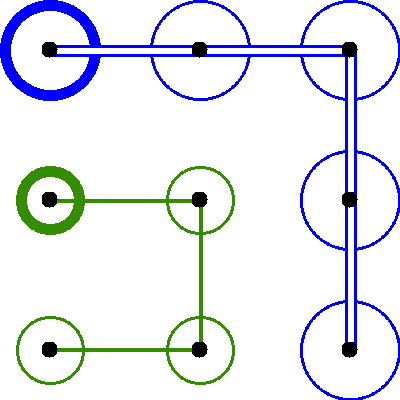}}
        \frame{\includegraphics[width=1cm]{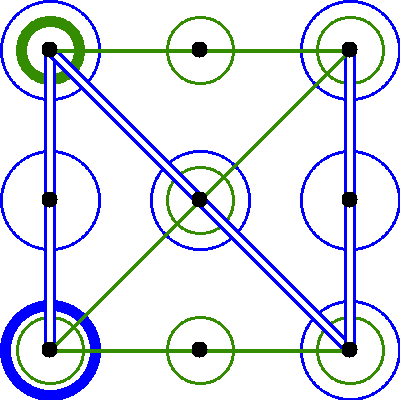}} \\
        \footnotesize{freq = 3} \\ \\
        \frame{\includegraphics[width=1cm]{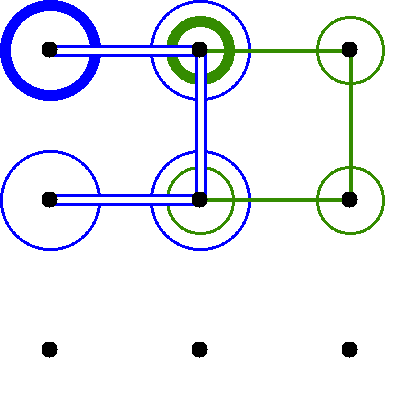}} 
        \frame{\includegraphics[width=1cm]{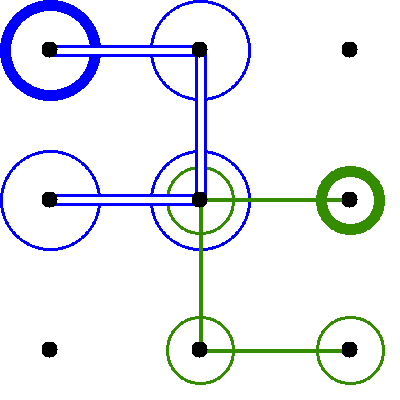}} 
        \frame{\includegraphics[width=1cm]{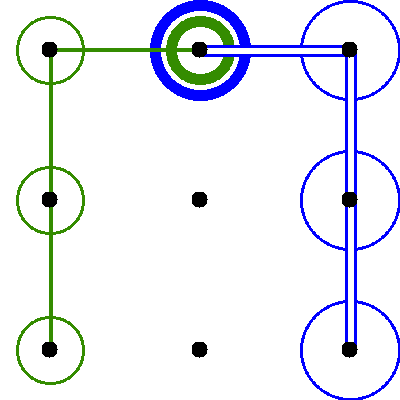}} \\
        \frame{\includegraphics[width=1cm]{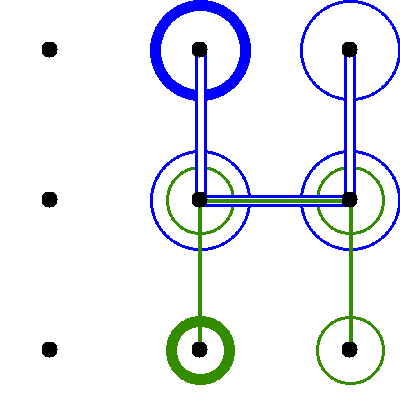}} 
        \frame{\includegraphics[width=1cm]{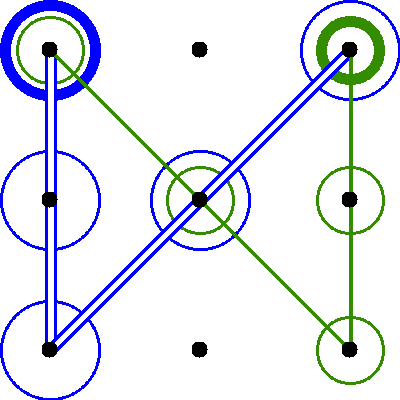}} 
        \frame{\includegraphics[width=1cm]{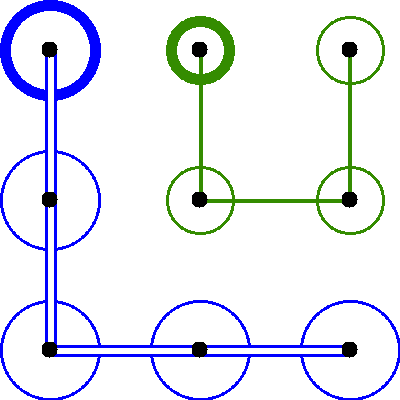}} \\
        \frame{\includegraphics[width=1cm]{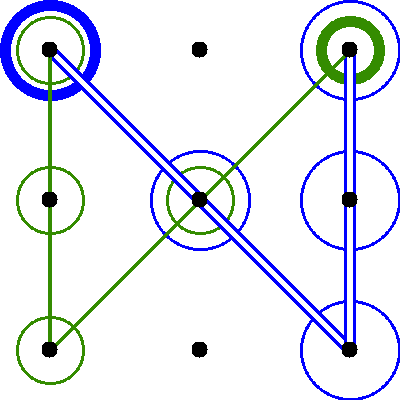}}
        \frame{\includegraphics[width=1cm]{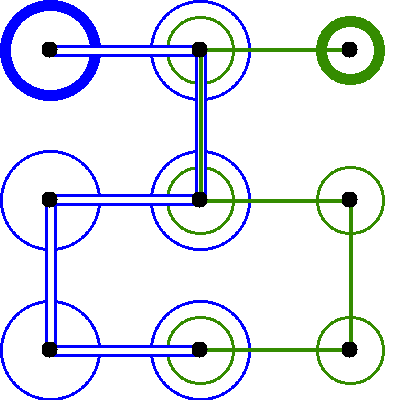}}
        \frame{\includegraphics[width=1cm]{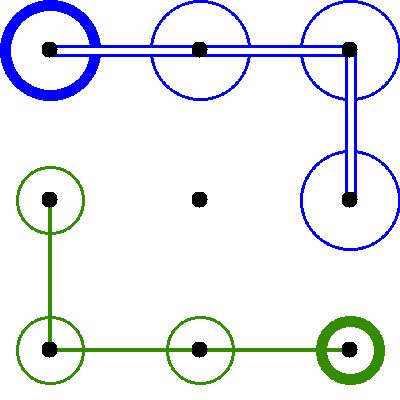}} \\
        \footnotesize{freq = 2} \\\\
%        \vspace{.8cm} \\
        \footnotesize{\emph{Remaining Double Patterns with}} \\
        \footnotesize{\emph{Single Occurrence Omitted}} \\
          }&
             
        % \makecell{ \\
        % \frame{\includegraphics[width=1cm]{dpatt-images/ideal/all/double-6-0143-5478.png}} 
        % \frame{\includegraphics[width=1cm]{dpatt-images/ideal/all/double-6-01258-03678.png}} 
        % \frame{\includegraphics[width=1cm]{dpatt-images/ideal/all/double-6-03678-01258.png}} \\
        % \footnotesize{freq = 6} \\ \\
        % \frame{\includegraphics[width=1cm]{dpatt-images/ideal/all/double-5-0341-7458.png}} \\
        % \footnotesize{freq = 5} \\ \\
        % \frame{\includegraphics[width=1cm]{dpatt-images/ideal/all/double-4-0143-1254.png}} \\
        % \frame{\includegraphics[width=1cm]{dpatt-images/ideal/all/double-4-0147-2147.png}} 
        % \frame{\includegraphics[width=1cm]{dpatt-images/ideal/all/double-4-0367-1258.png}} 
        % \frame{\includegraphics[width=1cm]{dpatt-images/ideal/all/double-4-03752-63158.png}} \\
        % \footnotesize{freq = 4} \\
        % }& 

             \makecell{ \\ 
          \frame{\includegraphics[width=1cm]{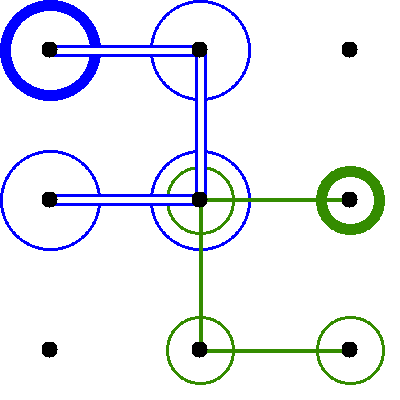}}
\ \ \          \frame{\includegraphics[width=1cm]{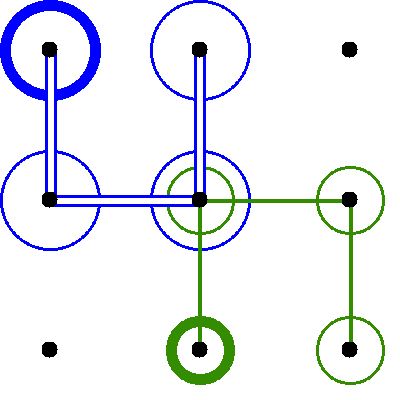}} \\
          \footnotesize{freq = 7} \ \ \  \footnotesize{freq = 5} \\\\
        \frame{\includegraphics[width=1cm]{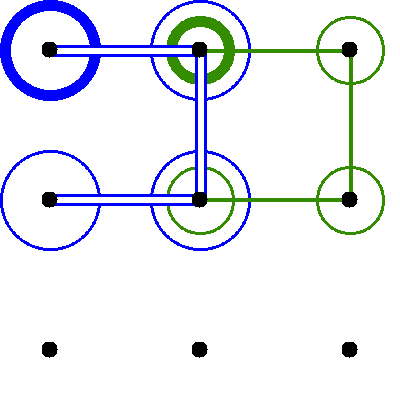}} 
        \frame{\includegraphics[width=1cm]{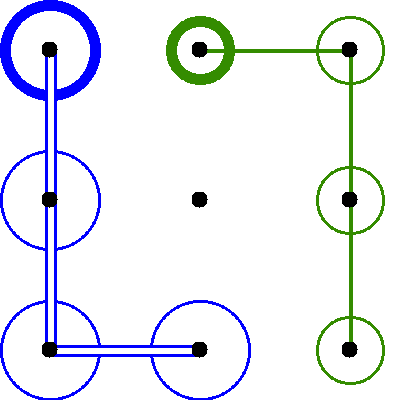}} 
        \frame{\includegraphics[width=1cm]{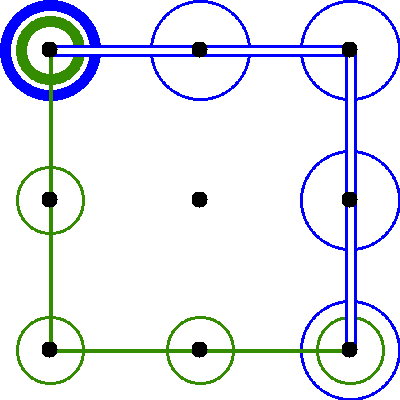}} \\
        \frame{\includegraphics[width=1cm]{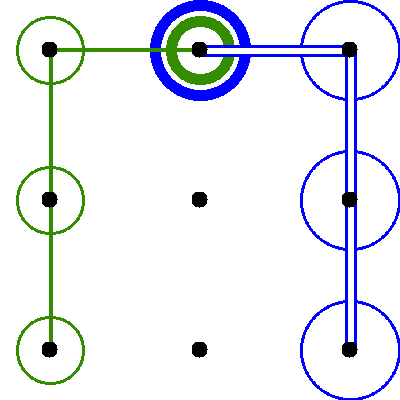}} 
        \frame{\includegraphics[width=1cm]{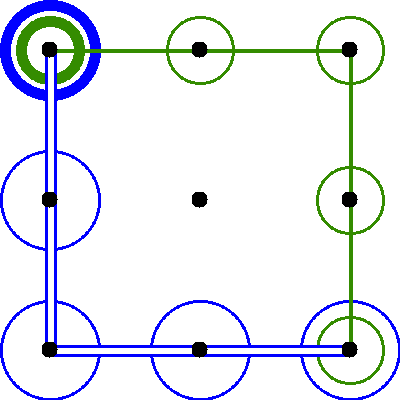}} 
        \frame{\includegraphics[width=1cm]{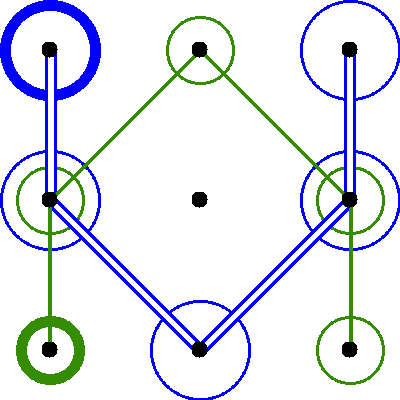}} \\    \frame{\includegraphics[width=1cm]{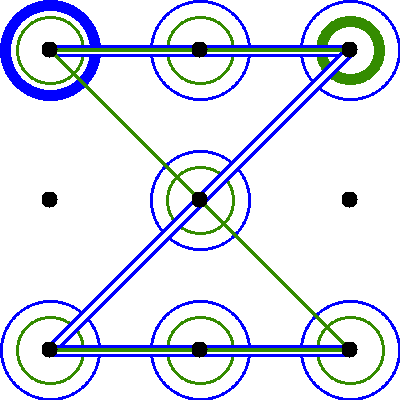}} 
        \frame{\includegraphics[width=1cm]{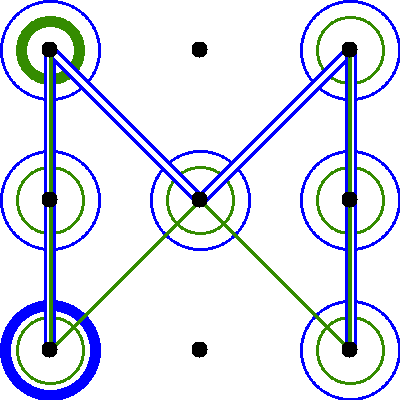}} 
        \frame{\includegraphics[width=1cm]{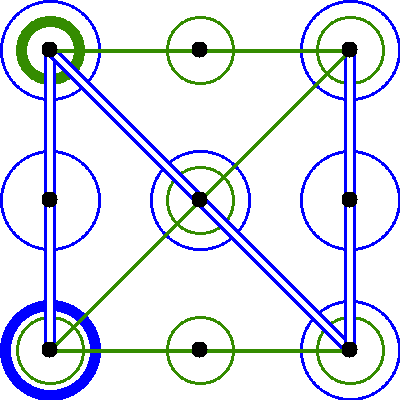}} \\
        \footnotesize{freq = 4} \\
        } \\
        % \hline
          \bottomrule
          \multicolumn{4}{l}{\em The blue pattern indicates the first pattern, and the green indicates the second pattern in the Double Pattern. Each contains a bold circle that denotes the starting point.}
	\end{tabular}}
      
\end{table*}

%%% Local Variables:
%%% mode: latex
%%% TeX-master: "../main"
%%% End:

%%% Local Variables:
%%% mode: latex
%%% TeX-master: "main"
%%% End:

\begin{figure}[t]
  \centering
\begin{tabular}{c }
%\multicolumn{2}{c}{\textbf{Overall}}\\
\toprule
  {\em Start Point} \\
  \includegraphics[width=0.7\linewidth]{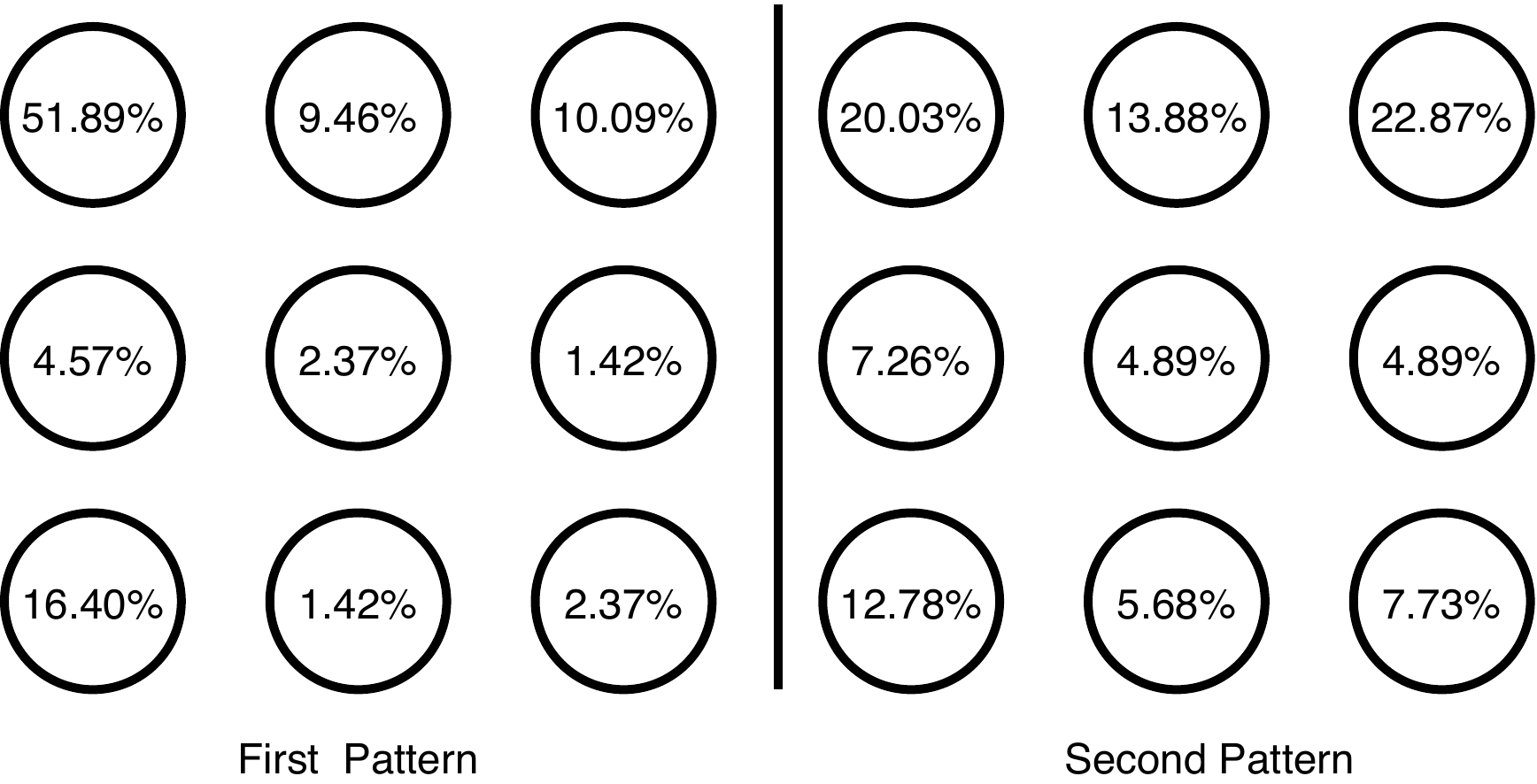} \\
  {\em End Point}\\
\includegraphics[width=0.7\linewidth]{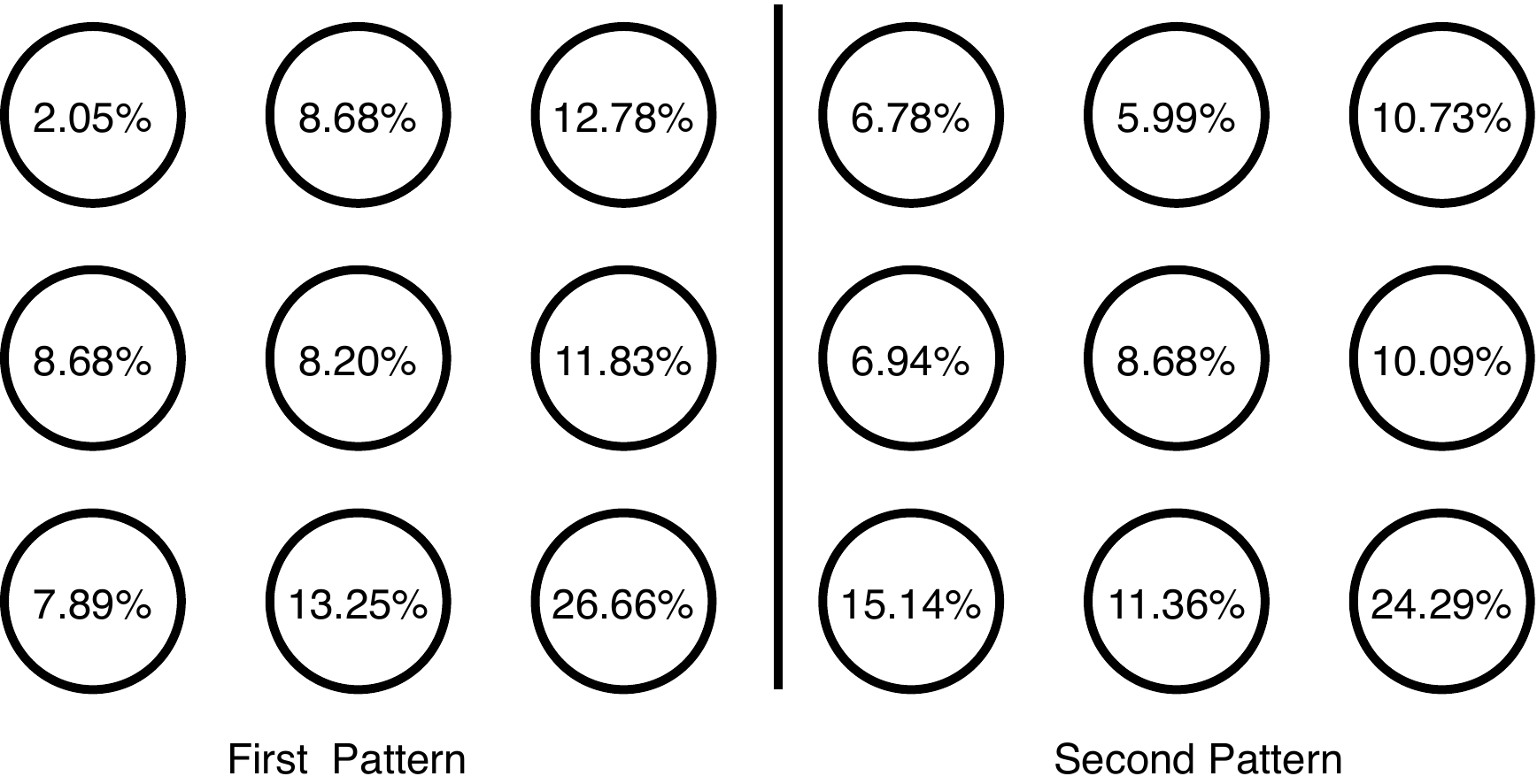}\\
\bottomrule

% \multicolumn{2}{c}{\makecell{\\\textbf{Control}}}\\
% \midrule
% {\em Start Point} & {\em End Point}\\
% \includegraphics[width=0.33\linewidth]{figures/startend/control-start.pdf} &
% \includegraphics[width=0.33\linewidth]{figures/startend/control-end.pdf}\\

% \multicolumn{2}{c}{\makecell{\\\textbf{Blocklist Both}}}\\
% \midrule
% {\em Start Point} & {\em End Point}\\
% \includegraphics[width=0.33\linewidth]{figures/startend/both-start.pdf} &
% \includegraphics[width=0.33\linewidth]{figures/startend/both-end.pdf}\\

% \multicolumn{2}{c}{\makecell{\\\textbf{Blocklist First}}}\\
% \midrule
% {\em Start Point} & {\em End Point}\\
% \includegraphics[width=0.33\linewidth]{figures/startend/first-start.pdf} &
% \includegraphics[width=0.33\linewidth]{figures/startend/first-end.pdf}\\

% \multicolumn{2}{c}{\makecell{\\\textbf{Ideal Choice}}}\\
% \midrule
% {\em Start Point} & {\em End Point}\\
% \includegraphics[width=0.33\linewidth]{figures/startend/ideal-start.pdf} &
% \includegraphics[width=0.33\linewidth]{figures/startend/ideal-end.pdf}\\
\end{tabular}
\caption{Frequence of Start/End Choices Across Treatments.}% \reminder{Maybe only show the All? Kind of the same for all of them}}
\label{fig:start_end}
\end{figure}

%%% Local Variables:
%%% mode: latex
%%% TeX-master: "../main"
%%% End:

\section{Results}
\label{sec:results}

In this section we describe the results of our analysis of security and usability of Double Patterns. First, we describe the statistics of DPatt choice, including the frequency of various DPatts and features therein. We then offer a security analysis using guessability as a metric and compare DPatt with other mobile authentication options, such as 4-/6-digit PINs and Android patterns. Finally, we provide analysis of the usability based on the SUS responses, entry/recall rates, and qualitative responses.

\paragraph{Datasets}
As described previously, the survey applies three randomized treatments: a control treatmenet with no intervention; a blocklist first (BL-First) treatment, where the first pattern of a DPatt is blocklisted; and a blocklist both (BL-Both) treatment, where the combination of the two patterns in a DPatt is blocklisted.

%We also introduce a combination dataset, the {\em ideal choice} (or just {\em ideal}) data set that describes what participants select as their {\em first preference} DPatt before any intervention and prior to confirmation. These DPatt represent what participants might select in an ideal setting, where they do not have to confirm or change their choice based on a blocklists. A similar data set for 4-/6-digit PINs was derived by Merkart et al.~\cite{markert-20-pin-blocklist}.

We also compare the security of DPatt to 4-/6-digit PINs from Markert et al.~\cite{markert-20-pin-blocklist}, which were collected with similar methodologies, a collection of 3x3 Android patterns used in a survey~\cite{avivSurvey} and originally collected in Aviv et. al~\cite{aviv2015bigger}, Uellenbeck et al.~\cite{uellenbeck-13-pattern}, Loge et al.~\cite{loge-16-pattern-user-choice}, and von Zezchwitz et al.~\cite{zezchwitz2016onquant}.
% and 4x4 Android patterns from Aviv et al.~\cite{aviv2015bigger}.
Additionally, we make use of a 4-digit PIN dataset collected by Daniel Amitay~\cite{amitay-11-iphone-pins}, and a dataset of 6-digit PINs derived from the RockYou password breach~\cite{cubrilovic-09-rockyou}. Both datasets are used in similar ways by Wang et al.~\cite{wang-17-pin} and Bonneau et al.~\cite{bonneau2012birthday}.

%\paragraph{Qualitative Analysis Methods}

\subsection{Double Patterns Features}
\label{sec:results:stats}

Table~\ref{tab:freq2} reports the most frequent patterns in each treatment. The first pattern of a DPatt is indicated in {\em blue}, and the second pattern is indicated in {\em green}. The starting contact point of each individual pattern is differentiated in bold. Common DPatts tend to be symmetric in shape; such as a box or flipped S's. A second common theme is non-overlapping/singularly-overlapping segments where the individual patterns only share a single point or no points in common, for example, rotated $\sqcap$ or $\sqcup$ shapes.

Observing the most common individual patterns, compared to the 3x3 patterns reported in Aviv et al.~\cite{avivSurvey}: 90.69\% of the first patterns and 86.75\% of the second patterns were previously observed in the dataset. Similarities of individual patterns is further supported when looking at the common start and end contact points, as presented in Figure~\ref{fig:start_end}. As was the case in prior work, participants are likely to start in the upper left and end in the lower right. However, this effect is less evident for the second component pattern, where the preference is more spread across the top row. This suggests that selecting the second pattern, with the presence of the visual first pattern, does alter some of the choices by individuals, as evident in the lower percentage of second patterns previously observed in the prior work.

When comparing the length (the number of points used in a pattern), we find that there is a significant difference between the length of the first component pattern and second pattern ($U=181136.5,p<0.001$), where the first pattern is slightly longer than the second. This suggest that participants are ``fitting in'' their second pattern into the shape of the first, and likely using fewer contact points to do that. There were no observed statistical differences between the length of individual patterns or the combination of patterns in a DPatt between the treatments.

After DPatt selection, participants were asked to describe their strategy regarding their chosen pattern, as well as Likert agreement towards two questions: if the Double Pattern provides adequate security, and if it was difficult to choose an appropriate Double Pattern for unlocking a personal device.  Examining a 25\% sub-sample of users, we coded their responses to the open question, and each participant was assigned between one and three codes, depending on the depth of their response. Regarding strategy, the most frequently cited strategies include aspects of visual characteristics (59.3\%), memorability (50.7\%), personal familiarity (11.3\%), usability (10\%), and security (4\%). Of the 25\% sub-sample, only a small portion attributed their decision to random choice (5.3\%). This is supported by the obvious structure observed in the patterns in Figure~\ref{tab:freq2}.

%%% Local Variables:
%%% mode: latex
%%% TeX-master: "main"
%%% End:

\begin{table*}[ht]
  \caption{Perfect Knowledge Attacker Guessing Metrics (Avg.[Med.] of 500 randomized runs)}
  %Results based on the average/median of randomized downsampling to smallest data set size (209). Reported statistics include: the number of patterns/PINs contained within the set being attacked,; the $\beta$-success-rate at 5, 10, and 30 guesses; the minimum entropy of the most frequently occurring member within the set; the partial guessing entropy measured in bits for 5\%, 10\%, and 20\% of the set. }
    \label{tab:guess_down}
    \centering
    \resizebox{\linewidth}{!}{%
    \begin{tabular}{ r | c | c c c | c | c c c }
    
    & $n$  & $ \lambda_3$ & $\lambda_{10}$ & $\lambda_{30}$ & $H_{\infty}$ & $\widetilde{G}_{0.05}$ & $\widetilde{G}_{0.10}$ & $\widetilde{G}_{0.20}$ \\
    \toprule
     Control & 209 & 6.22\% [6.22\%] & 15.31\% [15.31\%] & 28.23\% [28.23\%] & 3.73 [3.73] & 3.93 [3.93] & 4.22 [4.22] & 5.12 [5.12] \\
    %\hline
%    ${}^\dagger$ Ideal Choice & 570 & 4.22\% [4.31\%] & 10.79\% [11.00\%] & 21.04\% [21.05\%] & 4.12 [4.24] & 4.43 [4.40] & 4.81 [4.78] & 6.15 [6.16] \\
    %\hline
    ${}^\dagger$ BL-First & 211 & 2.87\% [2.87\%] & 8.61\% [8.61\%] & 18.18\% [18.18\%] & 4.65 [4.65] & 4.76 [4.76] & 5.27 [5.27] & 6.45 [6.45] \\
    %\hline
    ${}^\dagger$ BL-Both & 214 & 3.83\% [3.83\%] & 10.53\% [10.53\%] & 20.33\% [20.57\%] & 4.24 [4.24] & 4.56 [4.56] & 4.88 [4.88] & 6.24 [6.21] \\
    \midrule
    ${}^\dagger$ 3x3 Patterns~\cite{uellenbeck-13-pattern,aviv2015bigger,loge-16-pattern-user-choice, zezchwitz2016onquant} & 4637 & 7.36\% [7.18\%] & 17.67\% [17.70\%] & 35.17\% [35.41\%] & 3.52 [3.55] & 3.69 [3.67] & 3.99 [4.03] & 4.85 [4.89] \\
    %\hline
%    ${}^\dagger$ 4x4 Patterns~\cite{aviv2015bigger} & 501 & 7.86\% [7.66\%] & 18.27\% [18.18\%] & 32.56\% [32.54\%] & 3.49 [3.55] & 3.59 [3.58] & 3.91 [3.96] & 4.79 [4.79] \\
    %\hline
    ${}^\dagger$ 4-digit PINs~\cite{markert-20-pin-blocklist} & 851 & 4.20\% [4.31\%] & 10.02\% [10.05\%] & 19.79\% [19.62\%] & 3.96 [3.96] & 4.45 [4.40] & 4.98 [4.92] & 6.29 [6.32] \\
    %\hline
    ${}^\dagger$ 6-digit PINs~\cite{markert-20-pin-blocklist}  & 369 & 6.65\% [6.70\%] & 10.93\% [11.00\%] & 20.50\% [20.57\%] & 3.15 [3.15] & 3.56 [3.59] & 4.69 [4.68] & 6.22 [6.22] \\
    \bottomrule
    \multicolumn{9}{c}{\footnotesize ${}^\dagger$ Random downsampling to the size of Control (209 Double Patterns).} \\
    \end{tabular}}
    
\end{table*}

%%% Local Variables:
%%% mode: latex
%%% TeX-master: "../main"
%%% End:

\subsection{Security}
\label{sec:res:security}
In this section, we discuss the evaluation of security of DPatts. We first outline the threat model, and then provide guessability analysis for two attacker variants, a perfect-knowledge and simulated attacker

\paragraph{Threat model.}  We make the following assumptions about the attacker in our threat model. First, the attacker is generic and not targeting a specific victim. A targeted attacker may have additional information about the victims tendencies or have previous observations (e.g., shoulder surfing~\cite{schaub2012surfing,zezschwitz-15-pattern-shoulder-surfing,de-luca-14-shoulder-surfing,aviv-17-shoulder-surfing-baseline}), and thus, a generic attacker provides a lower bound for attacker performance. It also provides direct comparisons to other mobile authentication~\cite{avivSurvey,markert-20-pin-blocklist,aviv2015bigger,wang-17-pin,bonneau2012birthday}.

We consider two variations of the generic attacker: a {\em perfect knowledge} and a {\em simulated} attacker. A perfect knowledge attacker is an upper bound on the performance of a generic attacker, and assumes that the attacker knows the exact distribution of frequencies of authentication being guessed (the perfect knowledge), and thus always guesses the next most frequent pattern. A simulated attacker, however, has a set of training data of the authentication, and must use that information to guess a set of unknown authentication.

\begin{table}[t]
  \caption{Simulated Attacker Throttled Guessing Performance}
  \label{tab:simguess}
  \resizebox{\linewidth}{!}{
  \begin{tabular}{l | c c c | c c c c c c}
    & & \multicolumn{2}{c|}{{\bf Blocklist Hits}} & \multicolumn{2}{c}{{\bf 3 Guesses}} & \multicolumn{2}{c}{{\bf 10 Guesses}} & \multicolumn{2}{c}{{\bf 30 Guesses}} \\
    & {\bf n} & {\bf No.} & {\bf \%} & {\bf No.} & {\bf \%} & {\bf No.} & {\bf \%} & {\bf No.} & {\bf \%}\\
    \toprule
    Control & 209 & \multicolumn{2}{c|}{--} & 4 & 1.9\% & 9 & 4.3\% & 11 & 5.3\%\\
    BL-First & 211 & 70 & 33.2\% & 0 & 0\% & 2 & 1.0\% & 4 & 1.9\%\\
    BL-Both & 214 & 19 & 8.9\% & 0 & 0\% & 1 & 0.5\% & 2 & 0.9\%\\ 
    \midrule
    3x3 Patterns & 4637 &\multicolumn{2}{c|}{--} &  245 & 5.3\% & 556 & 13.0\% & 1089 & 23.6\%\\
    4-Digit PINs & 851  &\multicolumn{2}{c|}{--} & 27 & 3.2\% & 39 & 4.6\% & 65 & 7.6\% \\
    6-Digit PINs & 369  &\multicolumn{2}{c|}{--} & 19 & 8.1\% & 24 & 6.5\% & 33 & 9.0\% \\
  \end{tabular}}
\end{table}

\paragraph{Perfect knowledge attacker.} The primary results of the perfect knowledge attacker analysis is presented in Table~\ref{tab:guess_down}. We present the guessing statistics for DPatt, as well as comparisons to other mobile authentication for 3x3 Patterns~\cite{uellenbeck-13-pattern,aviv2015bigger,loge-16-pattern-user-choice, zezchwitz2016onquant} and 4-/6-digit PINs~\cite{markert-20-pin-blocklist}. As the data sets are of varied sizes, for a more fair comparisons we randomly down-sampled the larger data sets to 209 and report the average (and inset median) of 500 repetitions. We consider two metrics for a perfect knowledge attacker, as described by Bonneau et al.~\cite{bonneau2012science}.

First, for a throttled attacker who has a limited number of guesses, the {\em $\beta$-success-rate}, which describes the percentage of the dataset guessed after $\beta$ guesses. Reported as $\lambda_\beta$ in Table ~\ref{tab:guess_down}, one can observe that traditional 3x3 patterns (and also 4x4) have much worse (higher guessing percentages) than 4-/6-digit PINs; however, the DPatt improves the situation greatly. After 10 guesses, (control) DPatt perform more similarly to PINs, but after 30 guesses, the percentage of control DPatt guessed greatly increases. Using either blocklisting technique greatly degrades the attacker performance, where BL-First treatment produces an even stronger authentication then 4-/6-Digit PINs.

The $H_\infty$ statistic, which relates to a throttled attacker performance, describes how diverse (in bits of entropy) the most frequent authentication is in the data set. For example, this measures how common is the most common authentication, like ``password'' or ``1234,'' and how much benefit an attacker gains from just guessing the most common password.  While there is a small improvement for control DPatt as compared to other authentications, the blocklist treatments greatly decrease the commonality of the most common authentication. This suggests that minimal interaction in the selection process can lead to increased security in user choice.

The second metric, $\alpha$-guess-work, correlates with an unthrottled attacker that is unconstrained by the number of attempts to guess an authentication. Here we measure, in bits of entropy, how much ``work'' is required to guess an $\alpha$ fraction of the data set. Higher entropy describes more work for the attacker, and thus stronger authentication.

These results are indicated by $\widetilde{G}_{\alpha}$ in Table~\ref{tab:guess_down}. In all cases, we find that DPatt is more diverse (has a higher entropy) and thus more secure than traditional patterns. For guessing 20\% of the data, the control treatment DPatt is nearly 0.5 bits higher, and the BL-First treatment is nearly 1.5 bits higher. The security of DPatt, again, is more similar to that of 4-/6-digit PINs, and in some cases stronger.

\begin{figure*}
    \centering
    \begin{minipage}{0.49\linewidth}
    \includegraphics[width=\linewidth]{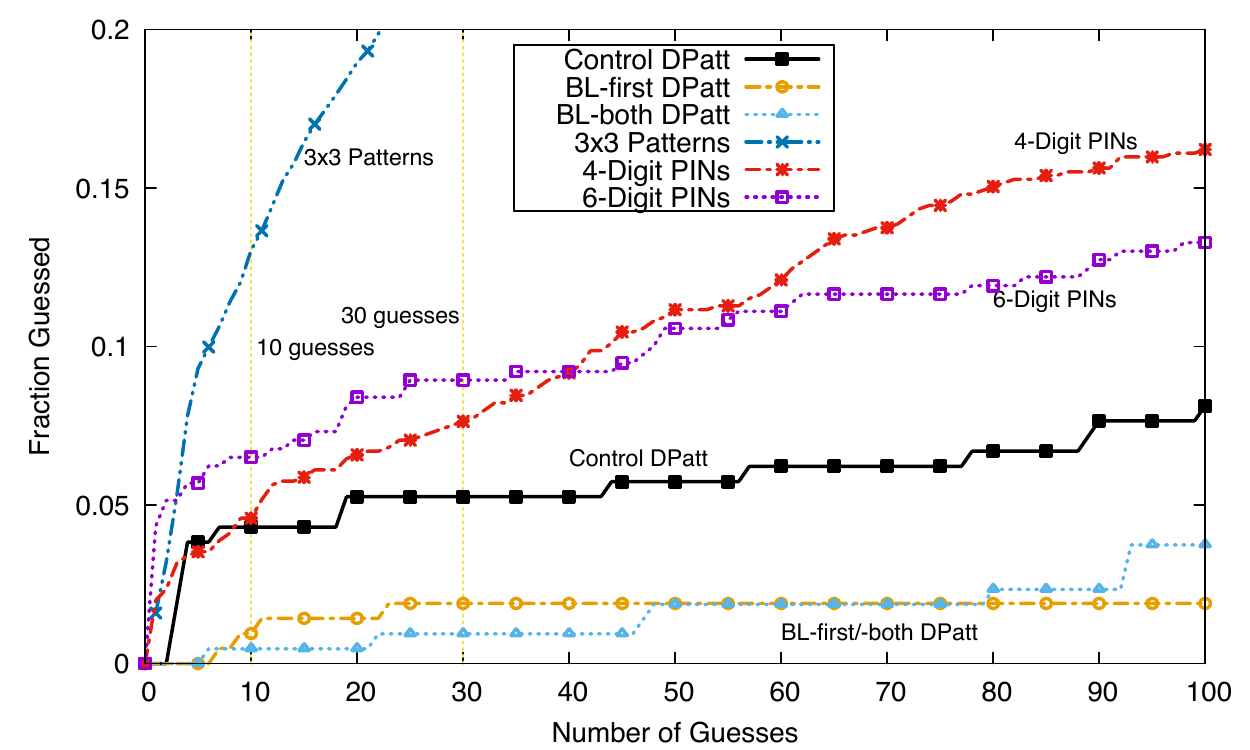}
    \caption{Simulated attacker on double pattern, first 100 guesses.}
    \label{fig:simguess:100}
    \end{minipage}
    \hfill
    \begin{minipage}{0.49\linewidth}
    \includegraphics[width=\linewidth]{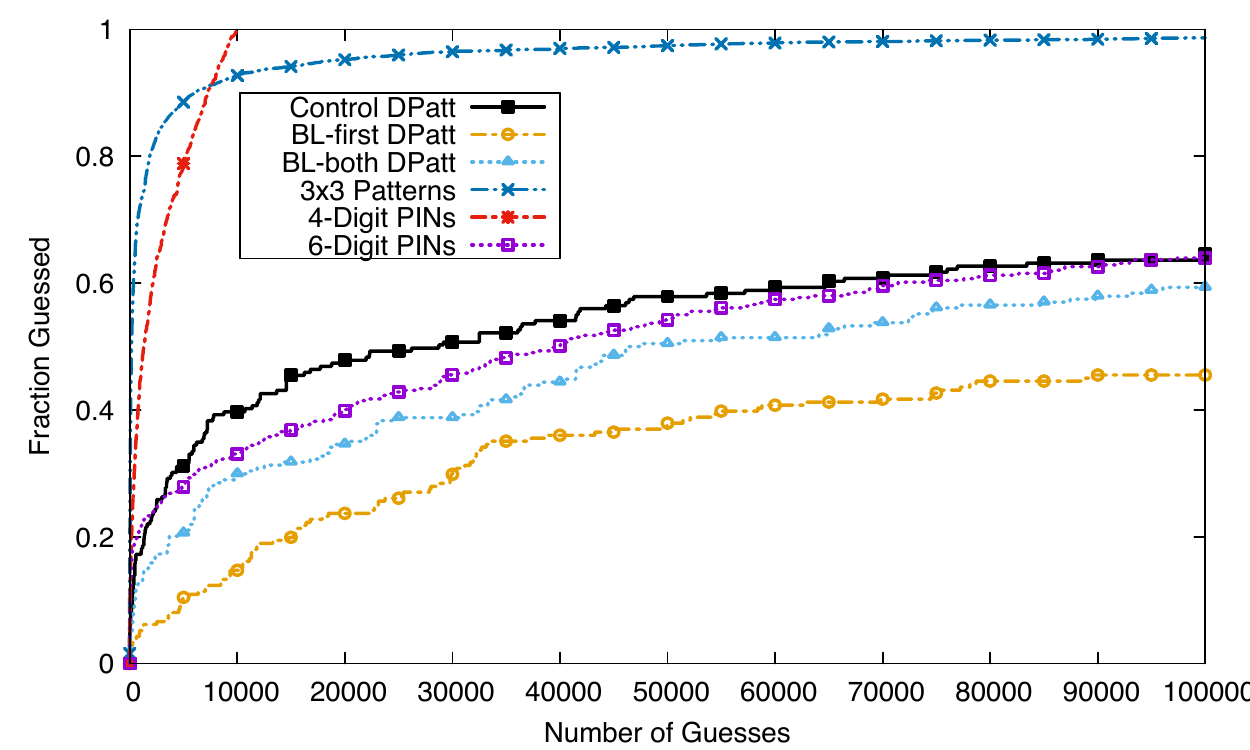}
    \caption{Simulated attacker on double pattern, first 100\,000 guesses.}
    \label{fig:simguess:100000}
    \end{minipage}
\end{figure*}

\begin{figure*}
    \centering
    \begin{minipage}{0.49\linewidth}
    \includegraphics[width=\linewidth]{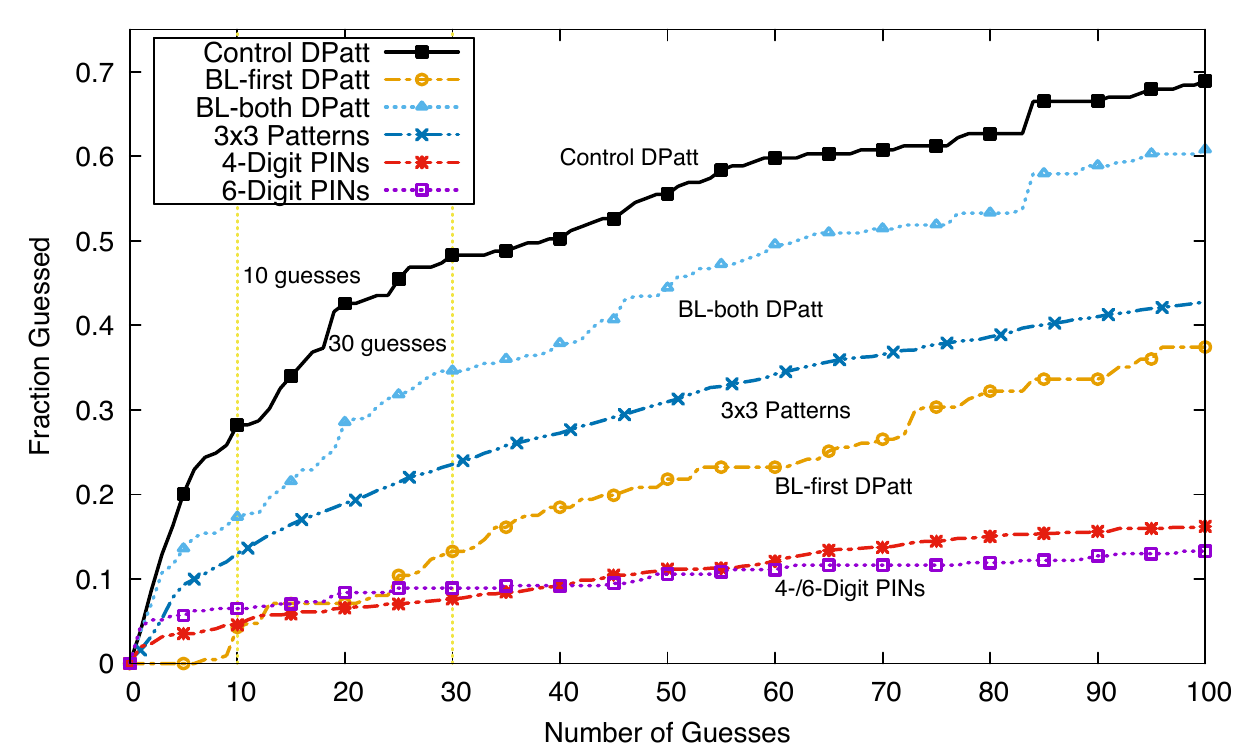}
    \caption{Simulated attacker on first pattern of double pattern.}
    \label{fig:simguess:fcomp}
    \end{minipage}
    \hfill
    \begin{minipage}{0.49\linewidth}
    \includegraphics[width=\linewidth]{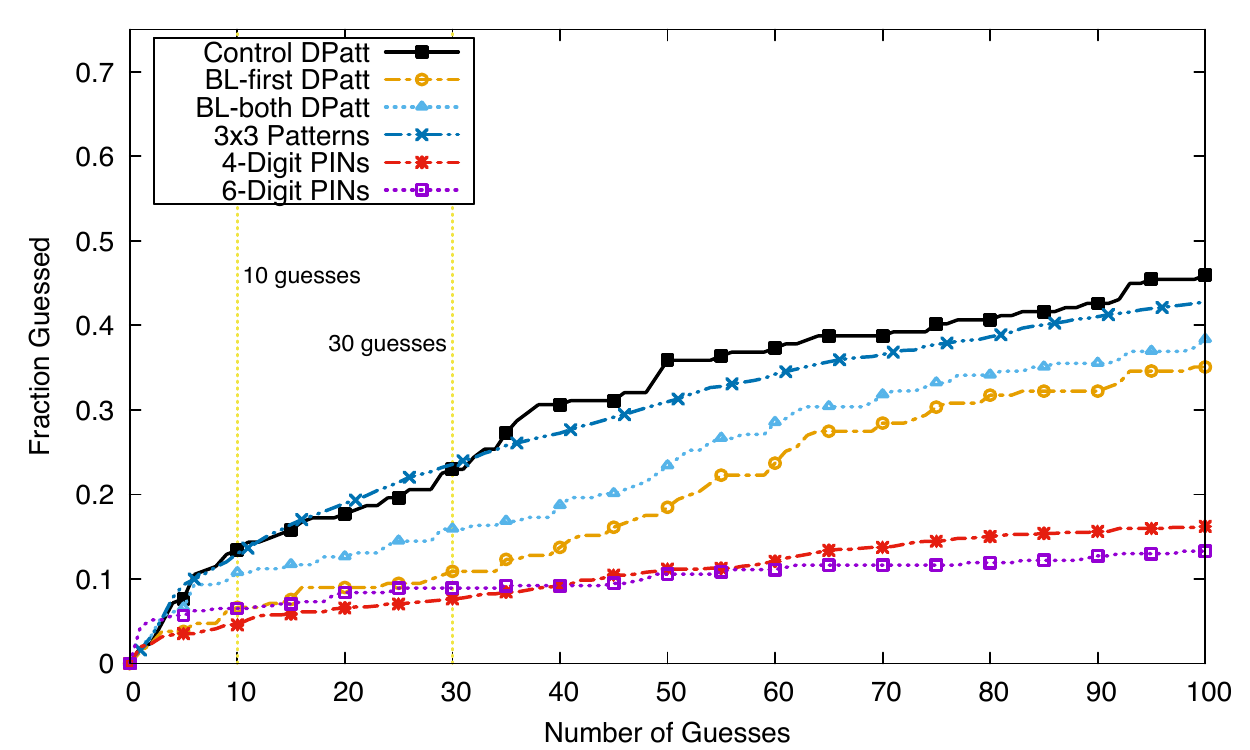}
    \caption{Simulated attacker on second pattern of double pattern.}
    \label{fig:simguess:scomp}
    \end{minipage}
\end{figure*}

\paragraph{Simulated attacker.} Recall that a simulated attacker must guess a set of unknown authentications based on a set of training data. One such way to model this situation includes cross-fold validation, where the data is divided into $n$ folds and the attacker trains on each of $n-1$ folds, guessing the remaining fold (the test set). As the DPatt data sets are not large enough for sufficient cross-fold validation, we take a different approach to generate a synthetic training set from traditional 3x3 patterns.

Comprising of 4,637 patterns, the simulated training set was constructed based on other published data of 3x3 patterns~\cite{uellenbeck-13-pattern,aviv2015bigger,loge-16-pattern-user-choice, zezchwitz2016onquant}. We transformed these into Double Patterns by matching each pattern with every other pattern, allowing for repetition of DPatts. We removed any invalid DPatts where the two patterns are the same. As an example: if there were 10 occurrences of `L' shaped patterns in the data set and 5 occurrences of `M' shaped patterns, the synthetic training set would have 50 `L-M' DPatts and 50 `M-L' DPatts, but no `M-M' nor `L-L' as these are invalid DPatts.

This method provided us with 21,421,974 DPatts, which we sorted based on frequency order. The simulated attacker then guessed DPatts from most frequent to least frequent in the synthetic data set.  More advanced techniques for ordering DPatts in the synthetic data set could be used, such as ordering completely by a Markov model, but we found through experimentation that simply guessing in frequency order was the attacker's best strategy where ties are broken by the Markov model. In the blocklist treatments, we assumed the attacker had knowledge of the blocklist, and thus avoided guessing disallowed DPatts.

In a world where DPatts are actively used, an attacker would instead train on known DPatts as used in the wild (or at least self-reported to be used). We could simulate such a scenario by performing a cross-validation simulated guesser, whereby we divide the data into $n$ groups, train on $n-1$ of them and guess the remaining. Unfortunately, the size of data is not sufficient to support this method. For example, with a standard cross-validation of 5 groups (or folds), the attacker would train $\sim$150 and only guess $\sim$50 DPatts, which is too small to potentially generalize. We instead opt for a simulated DPatt set. Future research on this topic, where additional DPatts were collected, could use this data as training to evaluate the security of newly collected DPatts. 

We used similar guessing techniques when comparing DPatt to 4-/6-digit PINs. We followed the same strategy outlined by Market et al.~\cite{markert-20-pin-blocklist} where they used the Amitay 4-digit data set~\cite{amitay-11-iphone-pins} and the RockYou 6-digit data set~\cite{cubrilovic-09-rockyou} to guess their sample of PINs. When comparing DPatt to 3x3 patterns, we used a cross-fold validation as there are no available secondary data sets to use of sufficient size, and followed the guessing methods outlined by Aviv et al.~\cite{aviv2015bigger}.

The main guessing results are presented in Figures~\ref{fig:simguess:100} and~\ref{fig:simguess:100000} and Table~\ref{tab:simguess}. Observe that DPatts, across all treatments, are more challenging for a simulated attacker to guess than other deployed authentication choices. At 30 guesses the attacker can only guess 5.3\% of the control treatment DPatts, compared to 23.6\% of traditional 3x3 patterns, 7.6\% of the 4-digit PINs, and 9.0\% of the 6-digit PINs. The disparity in strength between DPatt and other methods only increases with the implementation of blocklisting, where 1.9\% of the BL-First patterns and 0.9\% of the BL-Both patterns are discovered at 30 guesses. This suggests that significant security improvements could be gained from using DPatt even without a blocklist, but a blocklist would further enhance the security.

We also analyzed the individual patterns of a DPatt. In Figures~\ref{fig:simguess:fcomp} and~\ref{fig:simguess:scomp}, we perform simulated guessing of the first and second pattern (respectively) by guessing based on frequency order of the 3x3 data set. As before, we assume the attacker has knowledge of the blocklist. In the control treatment, the second component pattern is more difficult to guess than the first component (48.3\% vs. 22.9\% after 30 guesses), which suggest that participants choose more diverse second patterns to assist in visualizing a complete DPatt. Interestingly, the second component pattern of the control treatment is roughly as difficult to guess as traditional 3x3 patterns (23.6\% after 30 guesses). These results suggest that, without interventions, while participants select individual patterns of a DPatt that are no stronger (and often weaker) than selecting a single pattern, it is the combination of the two patterns in a DPatt that provides the added security.

%%% Local Variables:
%%% mode: latex
%%% TeX-master: "main"
%%% End:

\subsection{Usability}
\label{sec:results:usability}

In this section, we discuss the usability of DPatts based on the SUS scores, entry times, recall rates, response to security perception questions, and qualitative feedback. To code qualitative responses, we randomly selected a 25\% sub-sample of the responses (50 responses from each treatment). Two coders independently coded the responses and met to collaboratively code responses where coding differed.

%significance of the usability metrics we measured in regard to DPatt. We report results on selection/entry time considering attempts, Simple Usability Scale (SUS) feedback on our system, and willingness to adopt our interface.  

\paragraph{Entry/Selection time}
\begin{table*}[!!ht]
      \caption{Average {\em (stdev.)} Setup/Recall Time for Double Patterns (outliers removed using Tukey fencing)}
    \label{tab:attempts}
  \centering
  \resizebox{\linewidth}{!}{
    \begin{tabular}{ r | c c c | c c c c c }
    & \multicolumn{3}{c|}{\textbf{Setup}} & \multicolumn{5}{c}{\textbf{Recall}} \\
    \textbf{Treatment} & \textbf{Time} & \textbf{Attempts} & \textbf{Time/Attempt} & \textbf{Time} & \textbf{Attempts} & \textbf{Time/Attempt} & \textbf{Entry Time} & \textbf{Recall Rate} \\
    \toprule
    Control & 25.41s {\em (14.57s)} & 3.16 {\em (2.49)} & 5.26s {\em (2.41s)} & 4.74s {\em (2.80s)} & 1.36 {\em (0.86)} & 3.29s {\em (1.31s)} & 3.29s {\em (1.30s)} & 97.13\% \\
    BL-First & 35.50s {\em (25.28s)} & 4.45 {\em (3.58)} & 4.85s {\em (2.11s)} & 5.26s {\em (3.27s)} & 1.45 {\em (0.94)} & 3.51s {\em (1.45s)} & 3.54s {\em (1.40s)} & 94.79\% \\
    BL-Both & 23.44s {\em (12.74s)} & 3.47 {\em (2.62)} & 4.70s {\em (1.95s)} & 4.75s {\em (2.82s)} & 1.30 {\em (0.70)} & 3.21s {\em (1.23s)} & 3.19s {\em (1.19s)} & 97.20\% \\
    \bottomrule
    {\bf Total} & {\bf 27.14s {\em (16.93s)} } & {\bf  3.70 {\em (2.99)} } & {\bf  4.93s {\em (2.16s)} } & {\bf  4.94s {\em (3.01s)} } & {\bf  1.37 {\em (0.84)} } & {\bf  3.34s {\em (1.34s)} } & {\bf  3.35s {\em (1.31s)} } & {\bf  96.37\% } \\
\end{tabular}}
\end{table*}

%Add description in results
%%% Local Variables:
%%% mode: latex
%%% TeX-master: "../main"
%%% End:

Across all treatments, it took participants a mean time of 27.14s ({\em sd$=$16.93s}) to select a DPatt, taking an average of 3.70 attempts ({\em sd$=$2.99}) per participant, or 4.93s ({\em sd$=$2.16s}) per attempt. Recalling their DPatt is similar to an entry event, as in, participants do not need to complete the complex task of selection. When recalling their DPatt, participants spent an average of 4.94s ({\em sd$=$3.01s})  using 1.37 attempts ({\em sd$=$0.84}). Across all treatments the mean time per attempt was 3.34s ({\em sd$=$1.34s}), and the mean time per {\em correct} attempt was 3.35s ({\em sd$=$1.31s}). For comparison, related work has shown that Android patterns take on average 3.0s to enter and PIN's take 4.7s~\cite{harbach2014sa}, so DPatt adds only minimal time overhead to pattern entry. For a more detailed breakdown of selection and entry times, refer to Table~\ref{tab:attempts}. 

\paragraph{Perceptions of usability}
We use the System Usability Scale to measure participants perception of usability. Reported in Table~\ref{tab:sus}, across all treatments, an SUS score of 73.21 was reported, which is acceptable usability in the 60th percentile. However, when you break down the results based on current (or prior) Android pattern users, there is a much higher perception of usability. These participants provided an average SUS score of 78.27 which is in the 80th percentile for SUS. While there were dips in SUS due to blocklisting,  across all treatments Android pattern users rated DPatt more favorably.

\paragraph{Perceptions of security}
We asked participants to subjectively evaluate the security of DPatts in relation to existing method of authentication, using Likert agreement scale responses. We asked about the security of DPatt itself, and in comparison to original Android patterns, 4-digit PINs, 6-digit PINs, and alpha-numeric passwords. We also observed difference in responses of pattern users and non-patterns users ($U=31740.5, p< 0.001$).

Overall, participants responded positively to DPatt as a secure way to unlock their devices, 80\% either agreed or strongly agreed. When compared to the original pattern interface, 74\% either agreed or strongly agreed that DPatts were more secure. 82\% prior and current pattern users observed that the interface was more secure (agreed or strongly agreed), while 70\% of non-pattern users felt the same way. Similar trends were found in other results: only 55\% of non-pattern users felt DPatts were more secure than 4-digit PINs, but 76\% current pattern users did and 52\% felt it was even more secure than 6-digit PINs and alpha-numeric passwords (53\%). This suggests that current pattern users would feel confident in using DPatt due to security, and even non-pattern users have high security perceptions, up to 6-digit PINs. 

%pattern users average response is that they {\em agreed} that DPatt itself was secure, and that it was more secure than all of the suggested comparisons. Among non-pattern users they also {\em agreed} (on average) that DPatt was secure, however when comparing DPatt to existing interfaces they only {\em agreed} DPatt was stronger than the original Android patterns and 4-digit PIN's. Regarding 6-digit PIN's and alpha-numeric passwords, the overall sentiment was {\em Neither Agree Nor Disagree}.

We also collected Likert agreement responses regarding the perceived security of the DPatt selected by the participant: 83\% of our sub-sample agree (or strongly agree) that the DPatt they chose provided adequate security for unlocking their personal device.
%11.3\% {\em Neither Agree nor Disagree} with this sentiment, and only 6\% {\em disagreed} with $<1$\% {\em strongly disagreeing}.
With respect to selection difficulty, the results in this category were more evenly split: 32\% {\em strongly agreed} or {\em agreed} that it was difficult to choose an appropriate DPatt, 14\% {\em neither agreed nor disagreed}, and 54\% {\em strongly disagreed} or {\em disagreed} that it was difficult to choose. This suggest that most participants believe they are choosing secure DPatts and that it not difficult to do so. 

\paragraph{Willingness to adopt}
Our survey asked participants if they would, would not, or were unsure if they would utilize the Double Pattern they selected within the survey. Following this they were asked to expand on their choice in a free response form. Differentiated by previous pattern use and treatment, the results can be found in Table~\ref{tab:would_use}. Across all treatments, 42.3\% reported they would be comfortable using the DPatt they selected, 30.1\% reported they would not, and 27.6\% were unsure. Of the coded 25\% sub-sample, we found that the most frequently cited reason ($>$50\%) for non-utilization within the sub-sample was the notion that the participant's DPatt had been collected in the survey, so they would want to choose a new DPatt. We believe that this suggests DPatts found in the wild would likely be similar to those collected here, or at least more complex than those found in our survey results. 

Reflected in Table~\ref{tab:would_use_qual} in the Appendix, we found that the top three reasons participants would choose to utilize a DPatt as their authentication method is the memorability of the pattern they chose, the notion that they {\em like} the new interface itself, and the belief that DPatts themselves are secure, respectively. Regarding memorability, it makes sense that this is the top reason participants chose to utilize their DPatt, as we also asked participants to describe their strategies when choosing their DPatt during selection, and over half of the 25\% sub-sample reported making their DPatt memorable as an aspect of their strategy. 

Coinciding with memorability is the visual aspect of DPatts. Table~\ref{tab:freq2} portrays visual representations of the most frequently chosen DPatts within our survey. In addition, we examined participants' quantitative responses regarding DPatt selection. Reported in Table~\ref{tab:strat}, roughly 60\% of the sub-sample cites using visual aspects of DPatts in their selection strategy. Also detailed in Table~\ref{tab:strat} are participants' post pattern selection notions regarding their own creation strategy, as well as a self evaluation of their DPatt's security and how difficult it was for them to choose their DPatt.

%%% Local Variables:
%%% mode: latex
%%% TeX-master: "main"
%%% End:

\begin{table}[t]
    \caption{Simple Usability Scale sentiment.}
    \label{tab:sus} 
    \centering
    \resizebox{\linewidth}{!}{
    \begin{tabular}{c | c c c c| c}
    %\multicolumn{5}{c} {\textbf{Simple Usability Scale for Established Pattern Users vs Non-Pattern Users}} \\
    %\hline
    & \makecell[t]{\\ $\mathbf{n}$} & \makecell{\textbf{Num.}\\\textbf{Pat. Users}}  & \makecell{\textbf{SUS}\\\textbf{Pat. Users}} & \makecell{\textbf{SUS}\\\textbf{Non-Pat. Users}} & \makecell[t]{\\ \textbf{Combined}}\\
    \toprule
    Control & 209 & 57 & 78.55 & 70.90 & 72.99\\
    BL-First & 211 & 49 & 77.81 & 69.68 & 71.56 \\
    BL-Both & 214 & 56 & 78.39 & 73.84 & 75.04 \\
    \bottomrule
    {\bf Total} & {\bf 634} & {\bf 162} & {\bf 78.27} & {\bf 71.47} & {\bf 73.21} \\
    \end{tabular}}

    \end{table}

%%% Local Variables:
%%% mode: latex
%%% TeX-master: "../main"
%%% End:

\begin{figure}[t]
\centering
\includegraphics[width=\linewidth]{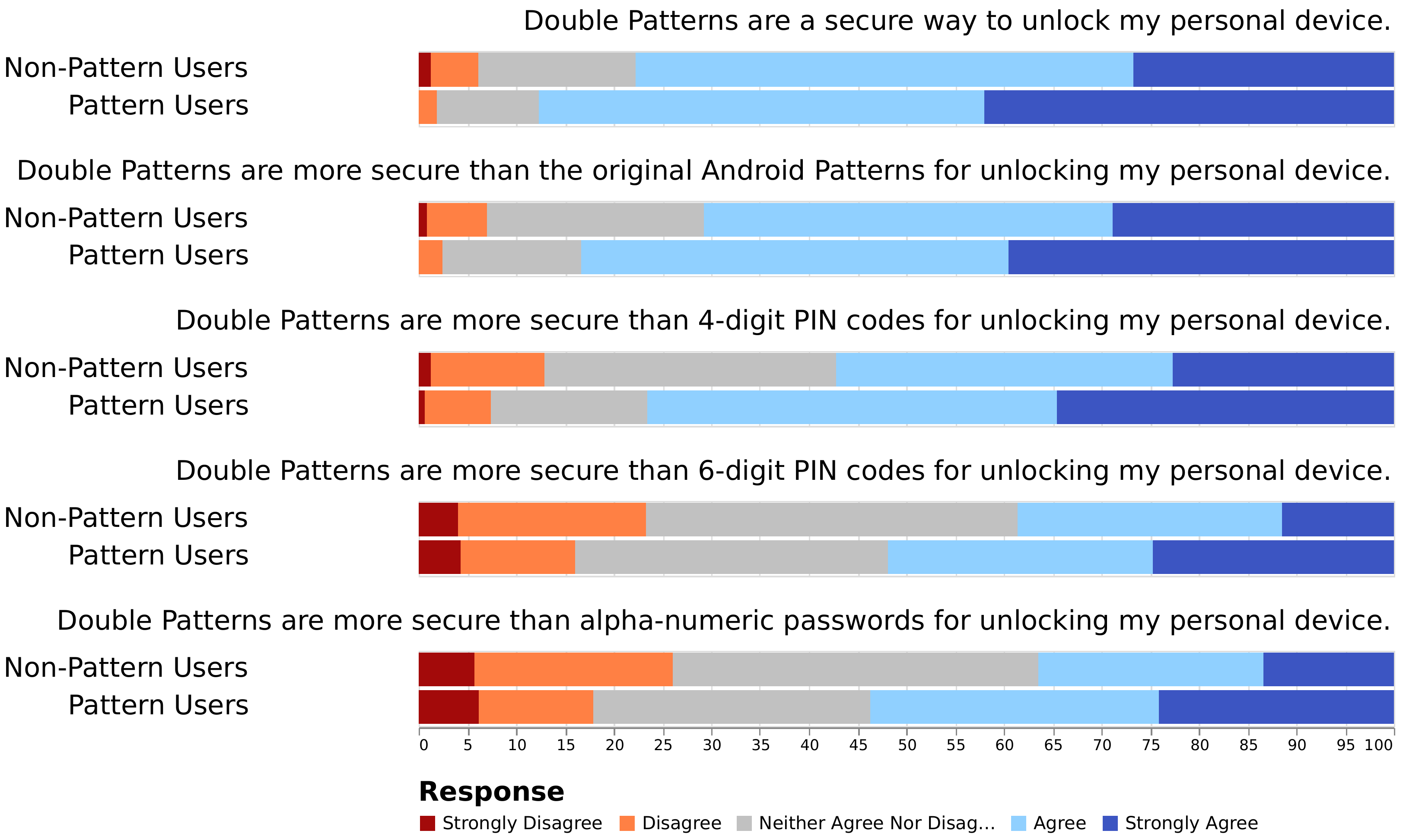}
\caption{Likert Results for Pattern Users Comparison.}
\label{fig:compare}
\end{figure}

\section{Discussion}
\label{sec:discussion}

Android pattern users continue to be a large cross-section of mobile device users, \url{~25\%} in this study, and there has not been a significant implementation change in Android patterns since initially deployed in 2008. While still preferred less than PINs, it is fair to assume that this stable user base will continue to prefer the graphical password interface that is unique to Android devices. However, without viable alternatives and extensions that provide increased security without degrading the user experience, current Android pattern users are less protected then their counterparts. Our results indicate that Double Patterns (DPatts) offer real potential as a natural extension to traditional Android patterns, that would be readily adopted and naturally increase security.

DPatt has strong usability. Participants in our study entered DPatts at roughly the same entry speed, less than a second slower (3.35s vs. 3.0s estimate in previous work). There was also high memorability $>94\%$ for a short-term study. In qualitative feedback, only two participants from our 25\% sub-sample (of about 150 participants) noted they were concerned with DPatt being cumbersome. Moreover, participants offered more than acceptable SUS scores, and more encouraging, participants that already use an Android pattern rated its usability in the 80-84th percentile. In fact, in the casual feedback to the study, a few participants noted that they were excited to see DPatts come to their device soon, expecting this to be a new feature of Android patterns.

DPatts also greatly increase the security of Android patterns without potentially frustrating user interaction. While blocklisting further improved DPatt, even the control case provides increased security more comparable to 4-/6-digit PINs.  Other proposals require re-selection of Android patterns, e.g., SysPal~\cite{cho2017syspal} or meters~\cite{song-15-pattern-psm,sun-14-pattern-psm,andriotis-14-pattern-psm}, potentially frustrating users away from their preferred choices, which they may reselect anyway if systems were non-enforcing. DPatt instead would be viewed as a new extension, more similar from going for 4- to 6-digit PINs, naturally encouraging users to extend their prior selected pattern in a new way that would increase security without the need of additional interventions. This is evident by the fact that the individual patterns of a DPatt that participants selected in this survey are no more secure (or perhaps less secure) than traditional Android patterns; it is the combination of two patterns that provides the security. 

Finally, while most participants in our survey believed DPatts are a secure way to unlock their personal device, current Android pattern users perceive DPatt as particularly secure, especially in comparison to other methods. This is a crucial view that suggests DPatts would be readily adopted if available, particularly to pattern users. Users would not be willing to change their authentication method to a system that they believe will harm them, and it is clear that DPatts provide a strong incentive for this group to upgrade their security while maintaining their preferred graphical password method. 

%%% Local Variables:
%%% mode: latex
%%% TeX-master: "main"
%%% End:

\section{Conclusion}
\label{sec:conclusion}

In this paper, we proposed using Double Patterns (DPatts) as an extension to Android patterns, whereby users enter two patterns, in sequence and super-imposed, as their unlock authentication. We conducted an online survey with $n=634$ participants selecting DPatts in three treatments: a control treatment, first pattern blocklist, and a full, DPatt blocklist.

We find, that across treatments, DPatts greatly increase the security compared to traditional Android patterns. A simulated attacker that must guess an unknown DPatt based on some training data, would only guess 5.3\% of the DPatts in the training set after 30 attempts as compared to 23.6\% of Android patterns. Blocklisting could be a viable option for further improving security, only 1.9\% and 0.9\% of DPatts in the first-pattern blocklist and full DPatt blocklist (respectively).

DPatts also provide minimal (if any) degradation in usability. Even in a short survey, participants recalled their DPatts at high rates ($>94\%$), and entry time is comparable with other current authentication methods, 3.35s. Observing current Android pattern users, this group had very high usability ratings as well as positive perceptions of the security of DPatts. As this is the group most likely to adopt DPatts (if deployed), this suggests that DPatts would be well received as a natural extension to Android patterns.

%%% Local Variables:
%%% mode: latex
%%% TeX-master: "main"
%%% End:

% \begin{acks}
\section*{Acknowledgments}
We would like to thank Daniel S. Roche at USNA for coordination and advice on this project, and we thank Harshvardhan Verma for assistance with qualitative coding. We also thank Maximillian Gola and Philipp Markert for feedback on the survey. This material is based upon work supported by the National Science Foundation under Grants No. 1845300. Any opinions, findings, and conclusions or recommendations expressed in this material are those of the author(s) and do not necessarily reflect the views of the National Science Foundation.
%\end{acks}

%\clearpage

% \bibliographystyle{ACM-Reference-Format}
\bibliographystyle{abbrvnat}
\bibliography{ref}
%\clearpage
\appendix
\appendixpage
%\section*{Appendix}

%\subsection{Preliminary Survey}
%\input{07-Prelim}

\section{Main Survey}
\label{app:survey}

 \newcounter{qcounter}
 
{\em Purpose of Study and Task Description:}
    We are conducting an academic survey about the use of Double Patterns in mobile authentication, and you will be asked to complete a survey that will ask you to generate a number of patterns under different conditions.
   You are being asked to participate in a research study focused on the effectiveness of using multiple patterns for mobile authentication on an Android device. Androids implement pattern locks rather than traditional security parameters, for example, numeric PINs or Alphanumeric Passcodes. Our research will focus on implementing an additional pattern lock as an increased security measure, and we are investigating the effectiveness of such a method.
   You will be asked to complete a short survey that requires you to generate a set of Android patterns under a security scenario, such as locking your device. Your eventual choices will be used in the final evaluation, as well as your responses to a set of security and usability questions.
    The expected completion time of the survey is 8-10 minutes, and no more than 1 hour. You will be compensated \$1.00 for your participation.
    \smallskip
    
    \noindent {\em Device Usage Questions}\\
    \noindent When referring to "mobile devices" throughout this survey, consider these to include smartphones and tablet computers. Traditional laptop computers, two-in-one computers, like the Microsoft Surface, or e-readers, like the Amazon Kindle, are not considered mobile devices for the purposes of this survey.
      
     \newcommand{\mychoice}[1]{{$\circ$} #1 \, }
    \begin{enumerate}
      \item How many mobile devices do you use regularly?\\
      \mychoice{0} \mychoice{1} \mychoice{2} \mychoice{3} \mychoice{4+}
    \item What brands of smartphone do you use for personal use? (Select all that apply)\\
    \mychoice{Apple}
    \mychoice{Samsung}
    \mychoice{LG}
    \mychoice{Motorola}
    \mychoice{Google/Pixel/Nexus}
    \mychoice{Huawei}
    \mychoice{ZTE}
    \mychoice{Other}

    \item What biometric method do you use most often to unlock your primary personal smartphone?\\
    \mychoice{Fingerprint}
    \mychoice{Face}
    \mychoice{Iris}
    \mychoice{Other Biometric}
    \mychoice{I do not use a biometric}
    \mychoice{I do not use a smartphone}
    \mychoice{Prefer Not to Say}
    
    \item {\em If choose biometric:} You have indicated that you use a biometric on your smartphone. Please answer the following question related to your response. 
    How do you unlock your primary personal smartphone when you reboot the device or if your biometric fails?\\
    \mychoice{Pattern Unlock}
\mychoice{4-Digit PIN}
\mychoice{6-Digit PIN}
\mychoice{PIN of other length}
\mychoice{Alphanumeric Password}
\mychoice{I use an unlock method not listed}
\mychoice{I do not use a smartphone}
\mychoice{Prefer Not to Say}
    \\\\
    {\em If did not choose biometric:}  You have indicated that you do not use a biometric on your smartphone. Please answer the following question related to your response.
What unlock method do you use on your primary personal smartphone?\\
    \mychoice{Pattern Unlock}
\mychoice{4-Digit PIN}
\mychoice{6-Digit PIN}
\mychoice{PIN of other length}
\mychoice{Alphanumeric Password}
\mychoice{I use an unlock method not listed}
\mychoice{I do not use a smartphone}
\mychoice{Prefer Not to Say}
   \setcounter{qcounter}{\value{enumi}}
\end{enumerate}

%\begin{wrapfigure}{r}{0.4\linewidth}
%\includegraphics[width=0.9\linewidth]{figures/interface/pattern_unlock.jpg}
%\end{wrapfigure} 
\noindent{\em What are Android Pattern Locks?}
Pattern Locks are used to unlock your smartphone, like a PIN. Patterns require you to "draw" shape that connects at least four of the contact points without lifting your finger or repeating a contact point. Displayed below is the Pattern Lock interface on a Samsung Android mobile device. \\
\\
\noindent{\em What are Double Pattern Locks?}
Double Pattern Locks are the same as Pattern Locks but require you to "draw" two shapes on the same 3x3 grid of contact points. The combination of the two patterns entered in the same order is now used to unlock your smartphone.

Each pattern in a Double Pattern is drawn the same way as before, but once you finish drawing your first pattern by lifting your finger, you then draw a second pattern. When drawing your second pattern, the first pattern will be displayed, and you may reuse contact points from your first pattern in drawing your second. However, you may not use your first pattern as your second pattern.

In this survey, we are exploring the possibility of using Double Pattern Locks as a new way to secure mobile devices. On the next page, you will have a chance to practice entering a Double Pattern before proceeding with the rest of this survey, where we will ask you to select your own Double Pattern that you would use to unlock your personal smartphone. \\
\\
\noindent{\em Practice:}
Practice entering a Double Pattern. (see Figure ~\ref{fig:doublepattern} for visual.)\\
%\end{wrapfigure} 
\\
\noindent{\em Instructions:}
For this survey, you will be asked to create a Double Pattern you would likely use for a personal device unlock, such as you would use on your smartphone.\\
%\begin{center}
%\includegraphics[width=0.05\linewidth]{figures/interface/unlock.png}    
%\end{center}\\
You will need to recall this Double Pattern later in the survey, so choose something that is secure and memorable as you may use on your personal device.\\
We ask that you DO NOT write down your patterns or use other aids to help you remember.\\
I understand that I should not write down my patterns or use other aids to assist in the survey. \mychoice{I understand}\\
I understand that I will be asked to create a Double Pattern for a personal device unlock. \mychoice{I understand}\\
\noindent{\em Selection} \\
Create a Double Pattern for a Personal Device Unlock. (See Figure~\ref{fig:doublepattern} for visual.)\\
\\
\noindent{\em Post Entry Questions:}
Thinking about the Double Pattern Lock you just chose:

\begin{enumerate}
\setcounter{enumi}{\value{qcounter}}
\item I feel that the Double Pattern I created provides adequate security for unlocking my personal device.\\
\mychoice{Strongly Agree}
\mychoice{Agree}
\mychoice{Neither Agree Nor Disagree}
\mychoice{Disagree}
\mychoice{Strongly Disagree}

\item It was difficult for me to select a Double Pattern that I would use to unlock my personal device.\\
\mychoice{Strongly Agree}
\mychoice{Agree}
\mychoice{Neither Agree Nor Disagree}
\mychoice{Disagree}
\mychoice{Strongly Disagree}

\item Everyone has a strategy when choosing their authentication, what was your strategy when choosing a Double Pattern? [open text]
    
    \setcounter{qcounter}{\value{enumi}}
\end{enumerate}

\noindent{\em Simple Usability Scale:} Select your agreement/disagreement with the following statements. Please note that the term "system" refers to Double Pattern Unlock. (Likert Response:
\mychoice{Strongly Agree}
\mychoice{Agree}
\mychoice{Neither Agree Nor Disagree}
\mychoice{Disagree}
\mychoice{Strongly Disagree})

\begin{enumerate}
\setcounter{enumi}{\value{qcounter}}
\item I think that I would like to use this system frequently.
\item I found the system unnecessarily complex.
\item I thought the system was easy to use.
\item I think that I would need the support of a technical person to be able person to be able to use this system. 
\item I thought there was too much inconsistency in this system.
\item I found the various functions in this system were well integrated.
\item I would imagine that most people would learn to use this system very quickly.
\item Select Agree as the answer to this question. (attention check)
\item I found this system very cumbersome to use.
\item I felt very confident using this system.
\item I needed to learn a lot of things before I could get going with this system.
\setcounter{qcounter}{\value{enumi}}
\end{enumerate}

\noindent{\em Recall Double Pattern:}
Recall the selecting Double Pattern. (See Figure ~\ref{fig:doublepattern} for visual.)

\noindent{\em Security Comparison:} Select your agreement/disagreement with the following statements.  (Likert Response:
\mychoice{Strongly Agree}
\mychoice{Agree}
\mychoice{Neither Agree Nor Disagree}
\mychoice{Disagree}
\mychoice{Strongly Disagree}) (Randomized order.)

\begin{enumerate}
\setcounter{enumi}{\value{qcounter}}

    \item Unlock patterns are more secure than 6-digit PIN codes for unlocking my primary smartphone.

\item Unlock patterns are more secure than 4-digit PIN codes for unlocking my primary smartphone.

\item Unlock patterns are more secure than alphanumeric passwords for unlocking my primary smartphone.

\item Unlock patterns are a secure way to unlock my primary smartphone.
\setcounter{qcounter}{\value{enumi}}

\end{enumerate}

\noindent{\em Use Double Pattern from Survey:}
\begin{enumerate}
\setcounter{enumi}{\value{qcounter}}
\item In a situation where your biometric fails or your mobile device reboots and you are utilizing a Double Pattern to unlock your personal mobile device, would you use the Double Pattern you selected in this survey, or would you select a different one?\\
\mychoice{Yes, I would use the Double Pattern I created here on my personal device.}\\
\mychoice{No, I would not use the Double Pattern I created here and instead create a new Double Pattern on my personal device.}\\
\mychoice{Unsure, I may or may not use the Double Pattern I created here on my personal device.}
\item \ [You have indicated that you would use / You have indicated that you are unsure if you / You have indicated that you would not use if you would use] the Double Pattern that you created in this survey on your personal mobile device. Please expand on why you [would / are unsure if you you / would not] use the Double Pattern you created here. [Open Text]
\setcounter{qcounter}{\value{enumi}}
\end{enumerate}
\noindent{\em Please Enter Your Demographic Information:}
\begin{enumerate}
\setcounter{enumi}{\value{qcounter}}
\item Select your age:
\mychoice{18-24}
\mychoice{25-29}
\mychoice{30-34}
\mychoice{35-39}
\mychoice{40-44}
\mychoice{45-49}
\mychoice{50-54}
\mychoice{54-59}
\mychoice{60-64}
\mychoice{65+}
\mychoice{Prefer Not to Say}

\item Select your gender
\mychoice{Female}
\mychoice{Male}
\mychoice{Non-Binary/Third Gender}
\mychoice{Not Described Here}
\mychoice{Prefer Not to Say}

\item What is your dominate hand?
\mychoice{Left Handed}
\mychoice{Right Handed}
\mychoice{Ambidextrous}
\mychoice{Prefer Not to Say}

\item Where you live is best described as
\mychoice{Urban}
\mychoice{Suburban}
\mychoice{Rural}
\mychoice{Prefer Not to Say}

\item What is the shape of a red ball?
\mychoice{Red}
\mychoice{Blue}
\mychoice{Square}
\mychoice{Round}
\mychoice{Prefer Not to Say}

\item What is the highest degree or level of school you have completed?
\mychoice{Some high school}
\mychoice{High school}
\mychoice{Some college}
\mychoice{Trade, technical, or vocational training}
\mychoice{Associate's Degree}
\mychoice{Bachelor's Degree}
\mychoice{Master's Degree}
\mychoice{Professional degree}
\mychoice{Doctorate}
\mychoice{Prefer Not to Say}

\item Which of the following best describes your educational background or job field?
\mychoice{I have an education in, or work in, the field of computer science, computer engineering or IT.}
\mychoice{I do not have an education in, nor do I work in, the field of computer science, computer engineering or IT.}
\mychoice{Prefer Not to Say}
\setcounter{enumi}{\value{qcounter}}
\end{enumerate}

\section{Blocklists}
\label{app:blocklist}
Patterns referenced by upper left contact point as 0, moving left to right counting, where the lower right contact point is 8.
\begin{itemize}
\item BL-First: (0.3.6.7.8), (0.3.6.7), (0.1.2.5.8), (0.3.6.4), (0.1.4.7), (0.1.2.5), (0.3.6.7.8.5.2), (0.4.8.5), (0.3.4.5), (0.4.8.7.6), (6.3.0.1.2), (0.1.2.4.6), (0.1.2.4.6.7.8), (2.5.8.7.6), (6.3.0.1), (0.4.8.5.2), (6.4.2.5.8), (0.3.4.1), (6.3.0.4) ,(1.4.7.8)
\item BL-Both: (0.3.6.7.8 2.5.8.7.6), (0.3.6.7 1.2.5.8), (0.3.6.7 2.5.8.7), (0.4.8.5 2.4.6.3), (0.4.8.7.6 2.4.6.7.8), (0.3.6.7.8 8.5.2.1.0), (0.1.2.5.8 0.3.6.7.8), (0.1.4.7 2.1.4.7), (0.3.6.7.8.5.2 2.5.8.7.6.3.0), (0.1.2.5.8 8.5.2.1.0), (0.3.6.7.8 0.1.2.5.8), (2.5.8.7.6 0.3.6.7.8), (6.3.0.1 8.5.2.1), (0.1.2.5 3.6.7.8), (0.3.4.1 1.4.5.2), (0.3.6.7.8.5.2 6.3.0.1.2.5.8), (0.1.2.4.6 6.7.8.4.0), (0.3.4.7.8 2.5.4.7.6), (5.4.7.6 3.4.7.8), (0.3.4.5 1.4.7.8), 
\end{itemize}
%%% Local Variables:
%%% mode: latex
%%% TeX-master: "main"
%%% End:

\section{Additional Tables and Figures}
\label{app:demo}
\begin{table}[h]
%  \caption{Additional Demographics of Participants}
%  \label{tab:background}
  \footnotesize
  \centering
\begin{tabular}{ r | c c c | c}
 & Control & BL-First & BL-Both & Total \\
\toprule
Tech & 62 & 66 & 55 & 183 \\
No-Tech & 140 & 139 & 154 & 433 \\
Prefer not to say & 7 & 6 & 5 & 18 \\
\midrule
Left-Handed & 24 & 16 & 31 & 71 \\
Right-Handed & 177 & 188 & 176 & 541 \\
Ambidextrous & 7 & 7 & 7 & 21 \\
Prefer not to say & 1 & 0 & 0 & 1 \\
\midrule
Rural & 34 & 32 & 36 & 102 \\
Suburban & 98 & 108 & 116 & 322 \\
Urban & 75 & 71 & 62 & 208 \\
  Prefer not to say & 2 & 0 & 0 & 2 \\
  \bottomrule
\textbf{Total} & {\bf 209} & {\bf 211} & {\bf 214} & \textbf{634} \\
\end{tabular}
\end{table}

%%% Local Variables:
%%% mode: latex
%%% TeX-master: "../main"
%%% End:

%\subsection{Device Usage}
%\input{tables/dev_num}

%\subsection{Frequently Utilized Component Patterns}
%\input{tables/patt_freq}

%\subsection{Blocklists}

%\section{Feelings and Sentiments}
\begin{table*}[ht]
\footnotesize
\caption{Code Book "Strategy" using 25\% Sub-Sample (50 per-Treatmet)}
    \label{tab:strat}
    \centering
    \begin{tabular}{c c p{4.8in}}
    \multicolumn{3}{c}{\bf Everyone has a strategy when choosing their authentication, what was your strategy when choosing a Double Pattern?} \\
    \toprule
    {\em Code} & {\em Frequency} & {\em Sample Quote} \\
    \midrule
    Memorability-memorable	&	76	&	"Choosing something that was memorable but not predictable to anyone that may try to unlock my phone."	\\
    Visual-shape	&	22	&	"I go for a shape I like and that is easy to remember."	\\
    Visual-letter	&	19	&	"My first name starts with the letter C so I drew a big C and a little C."	\\
    Choice-personal	&	17	&	"First letter of my father's first name and my mother's first name."	\\
    Visual-simple	&	17	&	"I wanted something simple enough to remember."	\\
    Usability-feel	&	10	&	"Something that felt natural to me."	\\
    Choice-random	&	8	&	"To make it as random as possible."	\\
    Visual-number	&	8	&	"I viewed the dots as a 1-9 keypad and entered memorable numbers."	\\
    Visual-unique	&	7	&	"One that would be hard to replicate in the correct order but easy for me to remember."	\\
    Visual-related	&	6	&	"I chose patterns with similar motions that would be easy to remember."	\\
    Visual-symmetry	&	5	&	"I use symmetrical patterns."	\\
    Security-secure	&	4	&	"I tried to make it not too complicated because I knew I'd have to remember it without writing it down or anything. But I tried to make it not too simple so that it felt secure enough."	\\
    Visual-reverse	&	3	&	"Using one shape and mirroring it."	\\
    Usability-physical	&	2	&	"Easy to do one handed."	\\
    Usability-usable	&	2	&	"I wanted something that I could remember \& would be easy to do with either hand." \\
    Choice-confident	&	1	&	"I just liked the pattern I chose."	\\
    Feeling-dislike	&	1	&	"I didn't have a strategy because I've never used this method and don't intend to."	\\
    Guessability-hard	&	1	&	"I picked something that couldn't easily be guessed and at the same time not to difficult to memorize."	\\
    Thinking	&	1	&	"Thinking."	\\
    Security-visual	&	1	&	"I wanted something both memorable to me but difficult to watch."	\\
    Usability-timely	&	1	&	"Tried to use something that I could remember and was quick."	\\
    Visual-repeat	&	1	&	"Repeat the pattern to remember it better."	\\
    Visual-subset	&	1	&	"Nothing really but I stayed within a smaller area." \\
     \bottomrule
    \multicolumn{3}{l}{${}^*$ Note that each quote can be assigned multiple codes.}
    \\ \\
    \end{tabular}
    \begin{tabular}{ r | c | c | c | c | c |}
    \multicolumn{6}{c}{\bf \makecell{I feel that the Double Pattern I created provides adequate \\ security for unlocking my personal device.}} \\
    \cline{2-6} 
    & \makecell{Strongly \\ Agree} & \makecell{ \\ Agree} & \makecell{ \\ NAND} & \makecell{ \\ Disagree} & \makecell{Strongly \\ Disagree} \\ 
    \cline{2-6} 
    & 40 & 83 & 17 & 9 & 1 \\
    \cline{2-6} 
    \end{tabular}
    \begin{tabular}{ r | c | c | c | c | c |}
    \multicolumn{6}{c}{\bf \makecell{It was difficult for me to select a Double Pattern that \\ I would use to unlock my personal device.}} \\
    \cline{2-6} 
    & \makecell{Strongly \\ Agree} & \makecell{ \\ Agree} & \makecell{ \\ NAND} & \makecell{ \\ Disagree} & \makecell{Strongly \\ Disagree} \\ 
    \cline{2-6} 
    & 24 & 24 & 21 & 46 & 35 \\ 
    \cline{2-6} 
    \end{tabular}
\end{table*}
\clearpage

%%% Local Variables:
%%% mode: latex
%%% TeX-master: "../main"
%%% End:

\begin{table*}[h]
\footnotesize
\caption{Code Book "Would use DPatt" using 25\% Sub-Sample (50 per-Treatment)}
    \label{tab:would_use_qual}
    \centering
    \resizebox{\linewidth}{!}{
    \begin{tabular}{c c c p{4.2in}}
    \multicolumn{4}{c}{\bf \makecell{You have indicated that you ({\em would use} | {\em would not use} | {\em are unsure if you would use}) the Double Pattern  you created in this survey on your \\ personal mobile device. Please expand on why you ({\em would use} | {\em would not use} | {\em are unsure if you would use}) the Double Pattern you created here.}} \\
    \toprule
    {\em Choice} & {\em Code} & {\em Frequency} & \multicolumn{1}{c}{{\em Sample Quote}} \\
    \midrule
    \bf Would Use && \bf 56 & \\
    &	Memorability-memorable	&	27	&	"It's easy to remember and is similar to my current single pattern but more secure."	\\
    &	Feeling-like	&	21	&	"I liked the idea and I would like to test it for several weeks." \\
    &	Security-secure	&	16	&	"I thought it added a good measure of safety that I would like."	\\
    &	Choice-confident	&	8	&	"I think I came up with good pattern."	\\
    &	Guessability-hard	&	8	&	"It would be hard to guess but easy for me to remember."	\\
    &	Choice-previous-use	&	3	&	"I already use the one half of the pattern and have for as long as I've had an Android."	\\
    &	Choice-different	&	2	&	"I think that it would be the only patterns that I could remember for sure. If I created different ones I would definitely forget about them."	\\
    &	Security-visual	&	2	&	"Using Secret number code I will create the pattern.so no one can know my pattern."	\\
    &	Usability-feel	&	2	&	"It's easy enough to remember and I like the design."	\\
    &	Usability-timely	&	2	&	"It seems secure and it easy to remember. It also seems like it will be fast to enter each time."	\\
    &	Usability-usable	&	2	&	"It was easy to remember. It was easy to use."	\\
    &	Visual-complex	&	2	&	"It is complex yet I can remember it."	\\
    &	Visual-shape	&	2	&	"I draw flags regularly. It is also difficult for a stranger to guess."	\\
    &	Visual-unique	&	2	&	"It seems unique and complicated enough to detour people unlike PINs."	\\
    &	Choice-personal	&	1	&	"It's a pattern I'm already familiar with."	\\
    &	Recall	&	1	&	"Because of your recall."	\\
    &	Visual-letter	&	1	&	"The Z pattern has always been my pattern."	\\
    &	Visual-simple	&	1	&	"It's easy for me to remember. And even though it's a simple shape it would take a few tries for someone who saw it to get right since the same shape can be achieved in many ways."	\\
        \bf Would Not Use && \bf 49 & \\
    &	Choice-no-reuse	&	26	&	"I would not use this one because it has been recorded on this survey."	\\
    &	Choice-complexity	&	9	&	"I used it here and would probably make it slightly more complex."	\\
    &	Choice-different	&	7	&	"I can think of something else easier but more secure for me to remember."	\\
    &	Security-unsecure	&	6	&	"I wouldn't use the same pattern twice for the same reason I don't reuse passwords; it's unsecure."	\\
    &	Memorability-unmemorable	&	5	&	"It was way too difficult to remember."	\\
    &	Visual-simple	&	4	&	"It was a little too simple. I only needed to remember the pattern for the survey so I didn't choose anything too complex."	\\
    &	Choice-personal	&	2	&	"I'd want something more original and personal to me."	\\
    &	Feeling-dislike	&	2	&	"I would use a PIN."	\\
    &	Guessability-easy	&	2	&	"I think everyone I know would automatically guess that I would use this symbol as my password just knowing my sense of humor."	\\
    &	Usability-cumbersome	&	2	&	"I have issues with memory. I wouldnt want to be stuck in an emergency."	\\
    &	Usability-rushed	&	2	&	"I would want to think about what pattern I would use for a longer time than is available during this survey."	\\
    &	Feeling-like	&	1	&	"I would not want to use a double pattern that I had used or someone knew about. I love the idea of using a double pattern."	\\
    &	Memorability-memorable	&	1	&	"I made it easy so I could remember it. But that's not good for security reasons."	\\
    &	Visual-number	&	1	&	"I think the one I created is pretty standard and generic. I would use the same idea of creating a visual number with the pattern."	\\
        \bf Unsure && \bf 45 & \\
        &	Choice-complexity	&	10	&	"I might choose a more complicated one."	\\
    &	Choice-no-reuse	&	9	&	"I would likely want to mix it up and use a different one that hasn't been previously shown to anyone including on the study. But also I liked the patterns I came up with."	\\
    &	Choice-different	&	5	&	"I would have more time to select a secure double pattern."	\\
    &	Choice-confident	&	5	&	"I would probably try the one that I created here."	\\
    &	Feeling-dislike	&	5	&	"PIN is easier to remember."	\\
    &	Security-unsecure	&	5	&	"I may make my pattern different than the one shown before. I may think of a new pattern that would make my phone more secure and safer."	\\
    &	Memorability-unmemorable	&	4	&	"I had some trouble remembering it exactly at times so I may do something more simple but I think I'd get it with time."	\\
    &	Memorability-memorable	&	4	&	"It felt a little too easy however easy to remember. I may want something a little more complex."	\\
    &	Usability-cumbersome	&	3	&	"I don't feel like a longer line or more dots will help. Also drawing longer lines can sometimes give you errors."	\\
    &	Guessability-easy	&	3	&	"I don't think my pattern was difficult enough. I would want to make something harder for someone to guess."	\\
    &	Choice-random	&	2	&	I might have a different first response when setting up a new double pattern so it could vary or I could use the same thing if it pops up in my head first.	\\
    &	Usability-feel	&	2	&	"I have my pin memorized by muscle memory so its probably easier than a double pattern."	\\
    &	Usability-rushed	&	1	&	"I would like to have more time to select a secure double pattern."	\\
    &	No-password	&	1	&	"I dont usually have a password on my phone. But considering it in the future maybe."	\\
    &	Visual-complex	&	1	&	"It depends. I may use an even simpler pattern. I have already shared my pattern here so I may change it to something else."	\\
    &	Feeling-like	&	1	&	"If I had another device I would. I use one I have used for a long time right now."	\\
    &	Don't-know	&	1	&	"I don't know."	\\
    \bottomrule
    \multicolumn{4}{l}{${}^*$ Note that each quote can be assigned multiple codes.}
    \end{tabular}}
    
\end{table*}
\clearpage

%Based on numbers you can compare descriptors
%25% of data collaboratively coded -- deconflicted 
%Report on how many responses we coded - 50 from each treatment
%Talk about how quotes could have multiple codes

%%% Local Variables:
%%% mode: latex
%%% TeX-master: "../main"
%%% End:

%\input{tables/comp_all}
%\subsection{Simple Usability by Treatment}
%\begin{figure}
%\centering
%\includegraphics[width=\linewidth]{figures/sus/sus_pos.PNG}
%\caption{Positive Sentiment Questions Regarding Usability}
%\label{fig:sus_pos}
%\end{figure}

%\begin{figure}
%\centering
%\includegraphics[width=\linewidth]{figures/sus/sus_neg.PNG}
%\caption{Negative Sentiment Questions Regarding Usability}
%\label{fig:sus_neg}
%\end{figure}

% \subsection{Device Comparison by Treatment}
% \begin{figure}
% \centering
% \includegraphics[width=\linewidth]{figures/compare/comp_all.PNG}
% \caption{Likert Results for All Users Comparison}
% \label{fig:comp_all}
% \end{figure}

%\clearpage
%%% Local Variables:
%%% mode: latex
%%% TeX-master: "main"
%%% End:

\end{document}